\documentclass[iop]{emulateapj}       

\newcommand{\about}{\mbox{$\sim$}}
\newcommand{\minus}{\mbox{$-$}}
\newcommand{\Lir}{\mbox{$L_{\rm IR}$}}                   
\newcommand{\nhtwo}{\mbox{$n_{\rm H_2}$}}         
\newcommand{\Smol}{\mbox{$\Sigma_{\rm mol}$}}  
\newcommand{\Tmb}{\mbox{$T_{\rm mb}$}}              
\newcommand{\Tb}{\mbox{$T_{\rm b}$}}                     
\newcommand{\Vlsr}{\mbox{$V_{\rm LSR}$}}             
\newcommand{\Tp}{\mbox{$T_{\rm p}$}}                     
\newcommand{\Idv}{\mbox{$I\Delta V$}}                     
\newcommand{\Ipdv}{\mbox{$I'\Delta V$}}                     
\newcommand{\taubar}{\mbox{$\bar{\tau}$}}                     
\newcommand{\Lsol}{\mbox{$L_\odot$}}
\newcommand{\Msol}{\mbox{$M_\odot$}}
\newcommand{\hr}{\mbox{$^{\rm h}$}}
\newcommand{\mn}{\mbox{$^{\rm m}$}}
\newcommand{\asec}{\mbox{$''$}}
\newcommand{\kms}{\mbox{km s$^{-1}$}}

\newcommand{\sqkpc}{\mbox{kpc$^{2}$}}
\newcommand{\perbeam}{\mbox{beam$^{-1}$}}
\newcommand{\persquarecm}{\mbox{cm$^{-2}$}}
\newcommand{\percubiccm}{\mbox{cm$^{-3}$}}
\newcommand{\peryr}{\mbox{yr$^{-1}$}}

\newcommand{\persquarepc}{\mbox{pc$^{-2}$}}

\newcommand{\persqkpc}{\mbox{kpc$^{-2}$}}

\newcommand{\unitofX}{\mbox{cm$^{-2}$ (K km s$^{-1}$)$^{-1}$}} 
\newcommand{\perkms}{\mbox{(km s$^{-1}$)$^{-1}$}}  

\newcommand{\uv}{\mbox{$u$--$v$}}
\newcommand{\Xco}{\mbox{$X_{\rm CO}$}}

\newcommand{\twelveCO}{\mbox{$^{12}$CO}}      
\newcommand{\thirteenCO}{\mbox{$^{13}$CO}}    
\newcommand{\CeighteenO}{\mbox{C$^{18}$O}}  
\newcommand{\twelveC}{\mbox{$^{12}$C}}            
\newcommand{\thirteenC}{\mbox{$^{13}$C}}          
\newcommand{\sixteenO}{\mbox{$^{16}$O}}            
\newcommand{\eighteenO}{\mbox{$^{18}$O}}         
\newcommand{\HCthreeN}{\mbox{HC$_{3}$N}}                 
\newcommand{\HNCO}{\mbox{HNCO(10$_{0,10}$--9$_{0,9}$)}}   
\newcommand{\HH}{\mbox{H$_2$}}                         
\newcommand{\water}{\mbox{H$_2$O}}                   
\newcommand{\ammonia}{\mbox{NH$_3$}}            
\newcommand{\nd}{\nodata}

\newcommand{\citest}[1]{\citeauthor*{#1}}
\newcommand{\citesp}[1]{(\citeauthor*{#1})}

\shorttitle{Cloud Complexes in the CMZ of NGC 253}
\shortauthors{SAKAMOTO et al.}

\slugcomment{Accepted for publication in {\it ApJ}}

\begin{document}
\title{Star-forming Cloud Complexes in the Central Molecular Zone of NGC 253}

\author{Kazushi Sakamoto\altaffilmark{1}, 
Rui-Qing Mao\altaffilmark{2},
Satoki Matsushita\altaffilmark{1,3},
Alison B. Peck\altaffilmark{3},
\\
Tsuyoshi Sawada\altaffilmark{3,4},
and
Martina C. Wiedner\altaffilmark{5}
}
\altaffiltext{1}{Academia Sinica, Institute of Astronomy and Astrophysics, 
P.O. Box 23-141, Taipei 10617, Taiwan}
\altaffiltext{2}{Purple Mountain Observatory, Chinese Academy of Sciences, Nanjing, 210 008, China}
\altaffiltext{3}{Joint ALMA Office, Alonso de C\'{o}rdova 3107, Vitacura - Santiago, Chile}
\altaffiltext{4}{National Astronomical Observatory of Japan, Tokyo 181-8588, Japan} 
\altaffiltext{5}{LERMA, Observatoire de Paris, CNRS, 61 Av. de l'Observatoire, 75014 Paris, France}

\begin{abstract}
We report 350 and 230 GHz observations of molecular gas and dust 
in the starburst nucleus of \objectname{NGC 253} at 20--40 pc (1\asec--2\asec) resolution.
The data contain CO(3--2), HCN(4--3), 
CO(2--1), \thirteenCO(2--1), \CeighteenO(2--1),
and continuum at 0.87 mm and 1.3 mm toward the central kiloparsec.
The CO(2--1) size of the galaxy's central molecular zone (CMZ) is measured to be
about 300 pc $\times$ 100 pc at the half maximum of intensity.
Five clumps of dense and warm gas stand out in the CMZ at arcsecond resolution,
and they are associated with compact radio sources due to recent massive star formation. 
They contribute one third of the CO emission in the central 300 pc
and have \twelveCO\ peak brightness temperatures around 50 K, 
molecular gas column densities on the order of $10^{4}$ \Msol\ \persquarepc,
gas masses on the order of $10^{7}$ \Msol\ in the size scale of 20 pc, 
volume-averaged gas densities of $\nhtwo \sim 4000$ \percubiccm, and
high HCN-to-CO ratios 
suggestive of higher fractions of dense gas than in the surrounding environment.
It is suggested that these are natal molecular cloud complexes of massive star formation.
The CMZ of NGC 253 is also compared with that of our Galaxy in CO(2--1) at the
same 20 pc resolution.
Their overall gas distributions are strikingly similar.
The five molecular cloud complexes appear to be akin to such molecular complexes as 
Sgr A, Sgr B2, Sgr C, and the $l=1.3\arcdeg$ cloud in the Galactic center.
On the other hand, the starburst CMZ in NGC 253 has
higher temperatures and higher surface (and presumably volume) densities
than its non-starburst cousin.
\end{abstract}


\section{Introduction } \label{s.intro}
NGC 253 is among the nearest and infrared-brightest starburst galaxies in the sky
and is therefore one of the best targets in the local universe for starburst studies.
Its active star formation is located in the central half kpc of this barred spiral galaxy 
at the distance of 3.5 Mpc \citep[1\asec\ = 17 pc;][]{Rekola05}.
The starburst luminosity of \about $3\times10^{10} \Lsol$ \citep{Telesco80} suggests
a star formation rate of \about5 \Msol\ \peryr\ according to the conversion from \Lir\ to
star formation rate by \citet{Kennicutt98}.  
The intensity of this starburst is approximately 20 \Msol\ \peryr\ \persqkpc\ in star formation
surface density, among the highest (and comparable to that in \objectname{M82}) in the sample of
nearby circumnuclear starbursts in \citet{Kennicutt98} when mergers are excluded.

Previous high-resolution observations of star forming interstellar medium in the nucleus
of NGC 253 showed that a quantity of molecular gas,  about a few $10^8$\Msol, in the starburst region is
warm, dense, and kinematically disturbed
(\citealt{Canzian88}; \citealt{Carlstrom90}; \citealt{Paglione95,Paglione04};
\citealt{Peng96}; \citealt{Sorai00}; \citealt{Garcia00}; \citealt{Ott05}; \citealt{Sakamoto06}, hereafter S06;
\citealt{Minh07}; \citealt{Knudsen07}).
However, the spatial resolutions of these observations, including our own \citesp{Sakamoto06}, 
were modest ($\gtrsim$3\arcsec)
and their frequency coverages limited ($\leq$ 230 GHz).

In this paper we report 1\asec--2\asec\ (20--40 pc) resolution imaging of the starburst nucleus
of NGC 253 in five
molecular lines and continuum emission in the 230 GHz and 350 GHz bands.
Several molecular cloud complexes are detected in the central 300 pc where
the starburst is most intense. 
The gas clumps are associated with known signs of massive star formation in the region,
suggesting them to be the natal clouds of, or a formative environment for, massive clusters.
The goals of this paper are to provide an updated overview of the starbursting central
molecular zone and to characterize the star-forming molecular cloud
complexes there on the basis of our new high-resolution multi-wavelength data.
The high resolution enabled us to directly constrain gas properties, e.g., 
temperature from the line brightness temperatures 
and volume-averaged gas density from the mass and size of the molecular clumps.
We also compare the center of NGC 253 with the center of Milky Way
using CO(2--1) data of the same linear resolution.
We point out morphological similarities as well as differences in the gas properties
between these starburst and non-starburst galactic centers.

\section{SMA Observations and Data Reduction \label{s.obsred} }
The observations were made with the Submillimeter Array \citep[SMA;][]{Ho04}\footnote{
The Submillimeter Array is a joint
project between the Smithsonian Astrophysical Observatory and the
Academia Sinica Institute of Astronomy and Astrophysics, and is
funded by the Smithsonian Institution and the Academia Sinica.
} as logged in Table \ref{t.obslog}.
We simultaneously observed three CO isotopologues, 
CO, \thirteenCO, and \CeighteenO,  in their $J$=2--1 transitions in the 230 GHz band.
We collected 5.7 hr of data toward 
$\alpha$=00\hr47\mn33\fs17,  $\delta$=\minus25\arcdeg17\arcmin17\farcs1 (J2000)
at the baseline lengths between 8 and 509 m from six nights of observations.
Our old data used in \citest{Sakamoto06} were not combined to these data 
in order to verify our previous results with an independent dataset.
The 345 GHz observations were made in two nights in an array configuration
called compact-north covering baseline lengths of 7--124 m.
The CO(3--2) and HCN(4--3) lines were observed for one night each. 
Three positions, the center given above and two positions 14\arcsec\ from it 
along the position angle of 60\arcdeg, were
observed. 
The resulting mosaic covers almost the same area as the 230 GHz data.
Both bands have a common flux scale tied to the primary calibrator Uranus \citep{Griffin93}.
We used quasars J0132\minus169 and J2258\minus279 for our gain calibration and
such bright sources as Saturn, Uranus, J1924\minus292, and 3C454.3 for our passband calibration. 
Pointing was checked during the tracks using J1924\minus292.

The data were reduced in the same manner  as in \citest{Sakamoto06} ---
initial calibration with the SMA version of MMA \citep{Scoville93}, imaging in MIRIAD \citep{Sault95}, and
data analysis in NRAO AIPS \citep{Bridle94}.
Special care was taken to ensure consistency in flux scale between tracks, by comparing
line and continuum intensities on common baselines. 
The uncertainty of flux calibration is 10\% and 15\% for 230 GHz 
and 345 GHz, respectively.
The line data were continuum-subtracted in the \uv\ domain by using continuum
made from the channels off the CO and HCN lines.
The only other line feature that we noticed in our passband was a weak \HNCO\ line
($f_{\rm rest}= 219.798$ GHz) between \thirteenCO(2--1) and \CeighteenO(2--1). 
The line was detected at 5--10 $\sigma$ in a few of channel maps but its total flux is 6\% of the \CeighteenO(2--1) flux
and its inclusion to our continuum affects our continuum flux by only 2\%.
Continuum maps were made by averaging both sidebands; their mean frequencies (wavelengths) are
225 GHz (1.33 mm) and 345 GHz (0.87 mm).  
All maps were corrected for the sensitivity pattern of the primary beam and mosaicking.
Velocities in this paper are with respect to the local standard of rest and are defined in the radio convention.

\begin{figure}
\epsscale{0.3}
\plotone{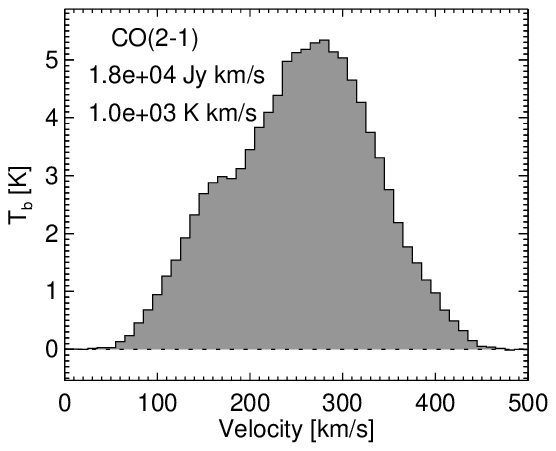} 
\plotone{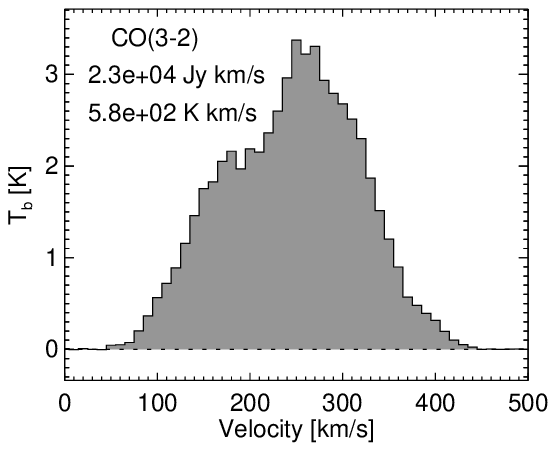} \\ 
\plotone{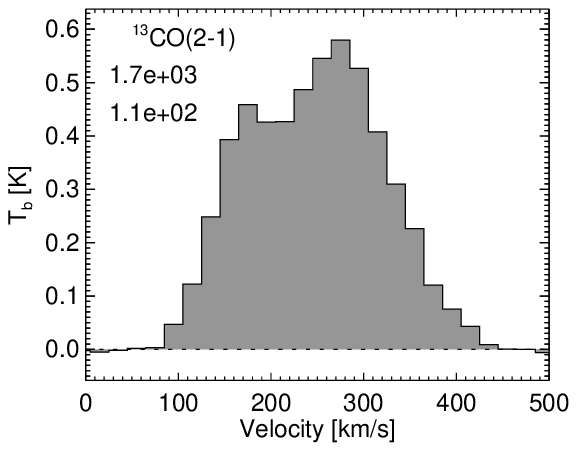}  
\plotone{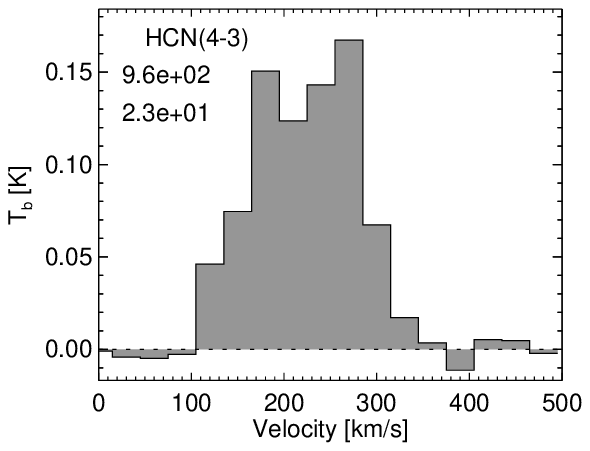} \\ 
\plotone{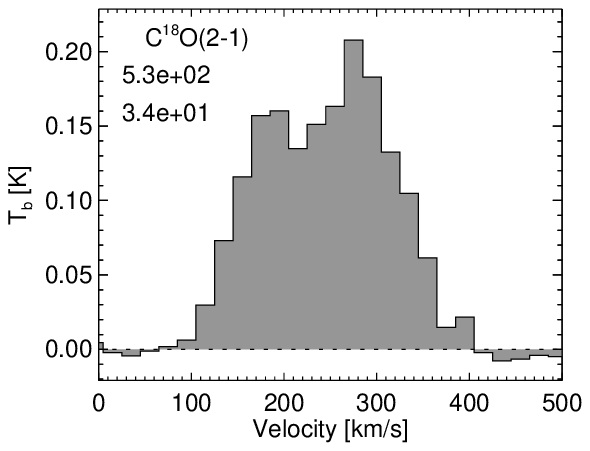}  
\plotone{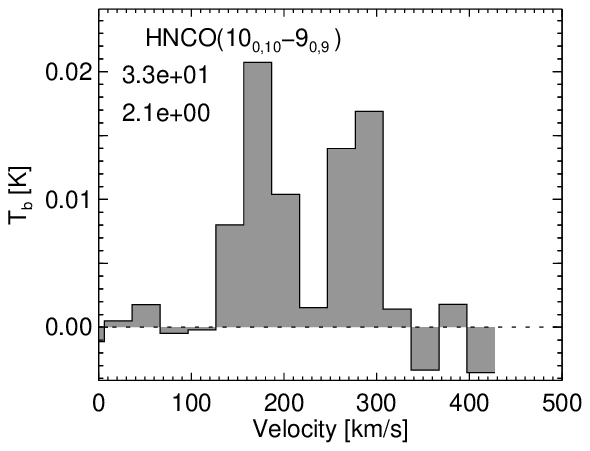} 
\epsscale{1.0}
\caption{Line emission at the center of NGC 253.
The spectra are taken from the SMA data cubes after 
correcting the cubes for the primary beam attenuation and convolving them to 20\arcsec\ resolution (FWHM).
The measured line fluxes are given in the legend of each panel in units of Jy \kms\ (second line) and K \kms\ (third line).
The continuum flux density at the center at the same resolution is 0.61 Jy \perbeam\ ($= 35$ mK) 
and 1.2 Jy \perbeam\ ($= 30$ mK) respectively at 1.3 mm and 0.87 mm. 
No correction for missing flux is made.
\label{f.spectra_center} }
\end{figure}

The parameters of our data sets are summarized in Table \ref{t.dataparam}.
Among them are fractions of single-dish flux as detected with our interferometric observations.
We detect most flux in the galactic center with slightly better recovery in the 230 GHz band.
Fig. \ref{f.spectra_center} shows the spectra and total fluxes of the lines detected at the center
of NGC 253.

\section{Results and Discussion \label{s.results}}
\subsection{Overview of the Central Molecular Zone}
\begin{figure}
\epsscale{0.57}
\plotone{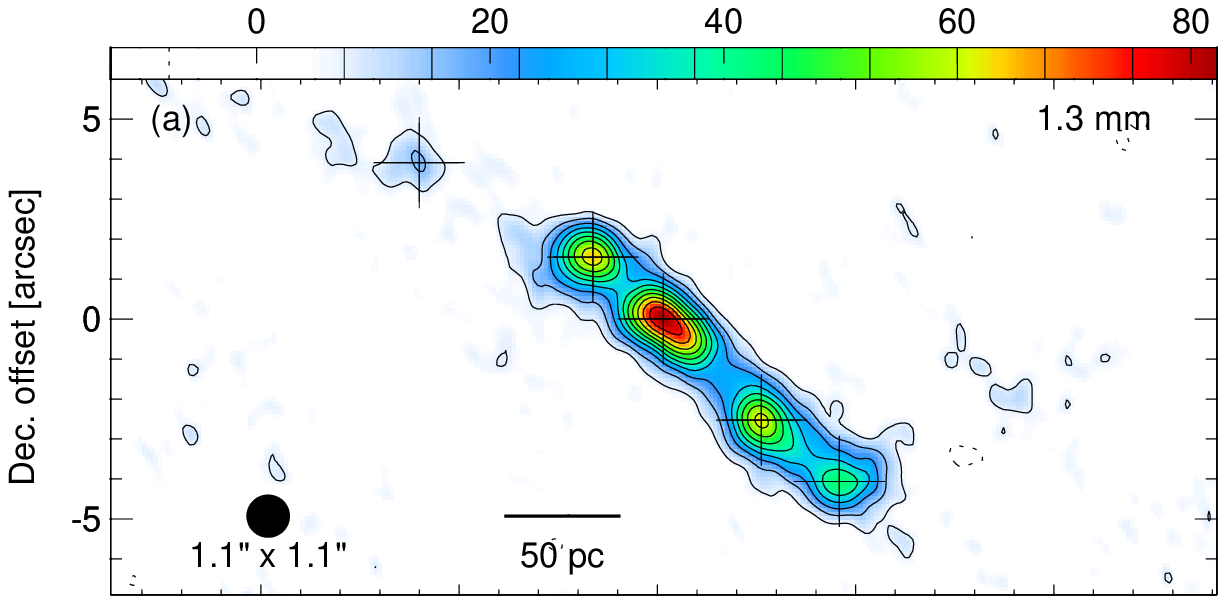} \\ 
\plotone{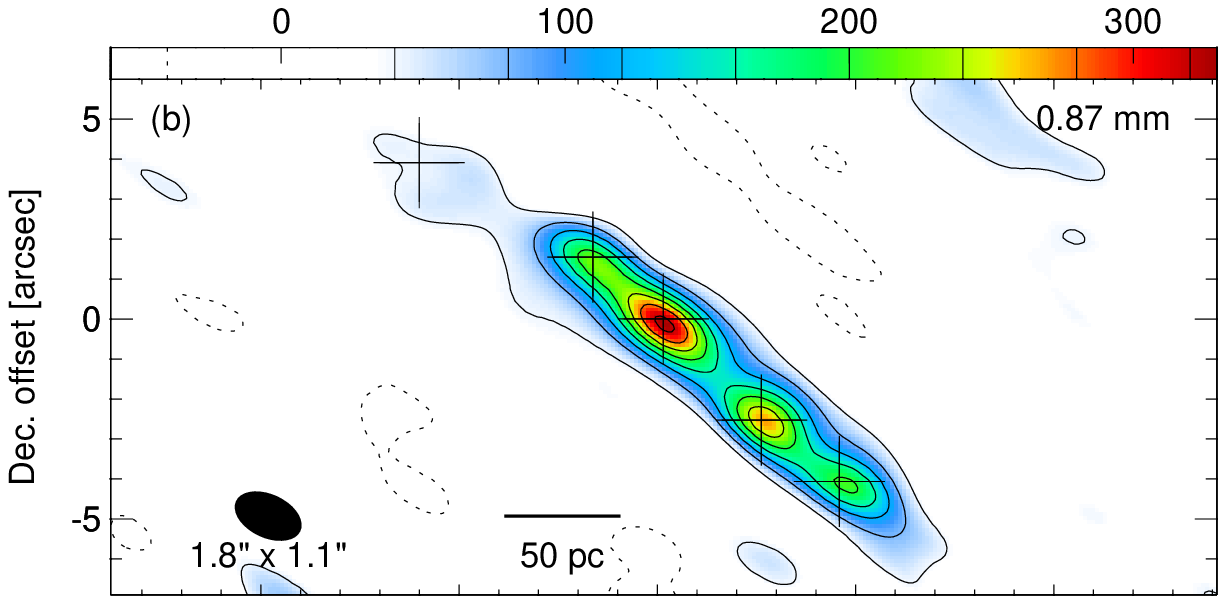} \\ 
\plotone{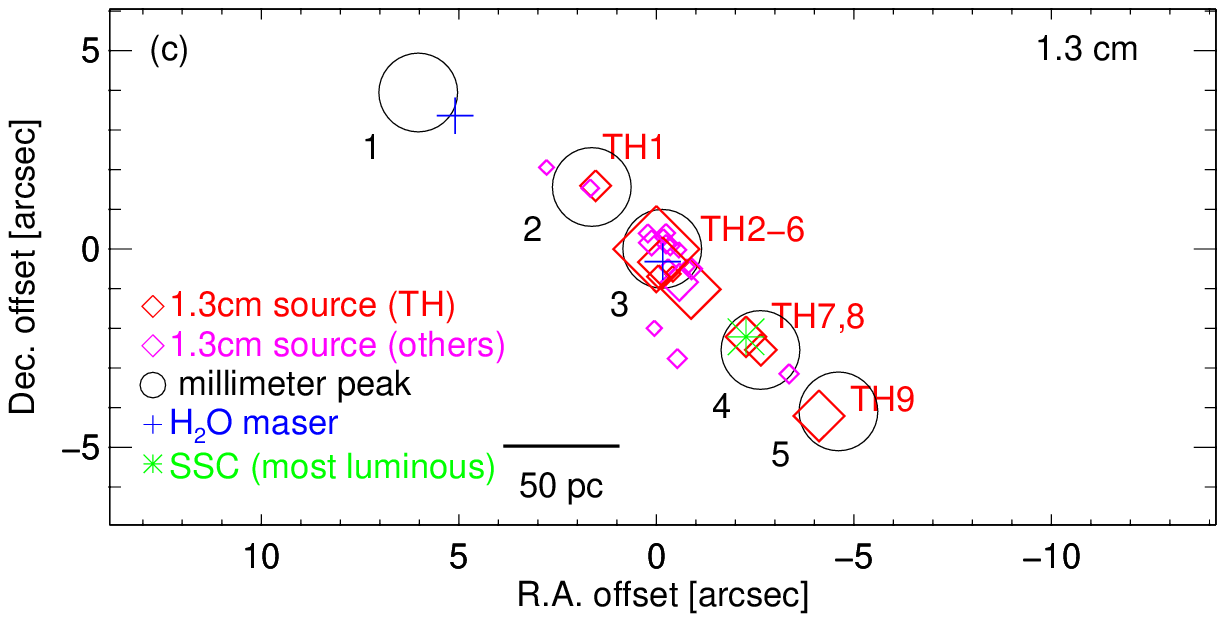} \\ 
\epsscale{1.0}
\caption{Radio continuum in the center of NGC 253. 
(a) 1.3 mm, (b) 0.87 mm, and (c) 1.3 cm and other sources for comparison.
In (a) and (b), contour steps are 7.5 mJy \perbeam\ ($=0.05 $ K $= 3\sigma$)
and 40 mJy \perbeam\ ($= 0.21 $ K $= 1.7\sigma$), respectively, and
peak intensities are 82 mJy \perbeam\ ($= 1.6$ K) 
and 0.32 Jy \perbeam\ ($= 1.7$ K), respectively.
The five brightest peaks of 1.3 mm continuum are marked with plus signs in (a) and (b) and with circles in (c).
Panel (c) shows the 1.3 cm compact sources listed in \citet[Table 6]{Ulvestad97} as diamond symbols, 
each of which has an area proportional to its flux density. 
The brighter sources found earlier by \Citet{Turner85} are indicated by red squares with TH$n$ names 
while others are shown in magenta. 
TH2 is the brightest source at 1.3 cm.
The green asterisk is the most IR-luminous star cluster in this region and is associated with TH7 
\citep{Keto99, Fernandez-Ontiveros09}.
Two blue plus signs are H$_2$O masers, one of which is at TH4 \citep{Henkel04,Hofner06}.
The origin of the offset coordinate in this and subsequent figures is 
$\alpha$=00\hr47\mn33\fs182,  $\delta$=\minus25\arcdeg17\arcmin17\farcs148 (J2000),
the position of TH2 measured by \citet{Lenc06}.
\label{f.cont} }
\end{figure}

\begin{figure}
\epsscale{0.57}
\plotone{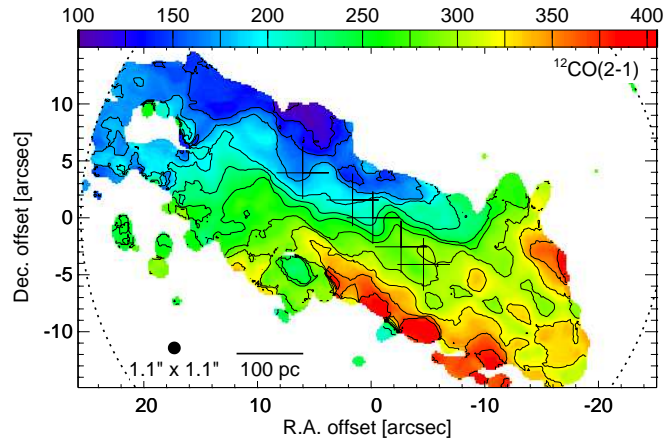}  
\epsscale{1.0}
\caption{Mean velocity map of \twelveCO(2--1) emission in the center of NGC 253.
Contours are in 25 \kms\ steps.
The five plus signs are at the positions of the 1.3 mm continuum peaks.
The approaching line of nodes of the galaxy is at the position angle of 51\arcdeg\ \citep{Jarrett03}.
\label{f.12co21.mom1} }
\end{figure}

Our line and continuum maps are shown in Figures  \ref{f.cont}, \ref{f.integ}, \ref{f.maxTb}, and \ref{f.12co21.mom1}.
The data clearly show the bar-like region of about 30\arcsec\ (500 pc) length
where the emission has been known to be strongest (e.g., \citest{Sakamoto06}).
Following the convention for our Galaxy, we call the region the central molecular zone (CMZ)
of NGC 253.
In the core of the CMZ are several peaks or clumps of molecular line and continuum emission.
In particular, there are five strong peaks in the 1.3 mm continuum map as listed in Table \ref{t.cont_peaks}. 
Each peak has a corresponding feature in our 0.87 mm map, 
in our \twelveCO(2--1) map,  and in many cases in other molecular line maps presented here.
The five peaks also have counterparts in existing observations at other wavelengths.
Some of these clumps have been seen in previous interferometric studies of molecular gas at lower resolutions
\citep[they are easier to see in more recent observations and discernible at least in ][]{Knudsen07,Minh07,Sakamoto06,Ott05,Paglione04,Garcia00,Peng96}. 
We put our main focus on these continuum and line peaks and describe their properties  in the following.
The millimeter peaks are hereafter referred to by their numbers in Fig. \ref{f.cont}(c) and Table \ref{t.cont_peaks}.

\subsection{Continuum Properties}
\subsubsection{Emission Mechanism inferred from Spectral Index}
The majority of continuum emission at around 1 mm is most likely thermal dust emission,
judging from the spectral index between 0.87 and 1.3 mm.
The flux density ratio of the 0.87 and 1.3 mm continua is  approximately 3 according to a data comparison
with matching range of \uv\ lengths. 
(The ratio $R_{0.87/1.3}$ is in the range of $3.2\pm 0.5$. See Table \ref{t.cont_peaks}.)
This steep increase of flux density toward shorter wavelengths, 
which was also seen between 3 and 1.3 mm (\citest{Sakamoto06} and references therein), 
is consistent with the dominance of thermal dust emission at $\lambda \lesssim 1$ mm.
The minimum fraction of dust emission is 0.8 at 0.87 mm for the flux density ratio
of $R_{0.87/1.3} = 3$, dust emissivity index of $\beta = 2$ \citep{Dunne01}, 
and a flat spectrum contribution from optically thin free-free emission.
The ratio at the nucleus, i.e., the peak No. 3, is $R_{0.87/1.3} =3.0$ and is not distinctly 
different from the ratio at the other peaks.

\subsubsection{Mass and Column Density from Continuum \label{s.continuum.mass}}
The peak column density of gas at the continuum peaks
is estimated to be (1--4) $\times 10^{4}$ \Msol\ \persquarepc\ 
from the 1.3 mm emission and  to be (0.2--2) $\times 10^{4}$ \Msol\ \persquarepc\ 
from 0.87 mm (Table \ref{t.cont_peaks}).
These estimates use the mass opacity coefficient of \citet{Hildebrand83}, 
$\beta=2$,
gas-to-dust mass ratio of 100,
and a dust temperature of 50 K.
The peak column densities from 1.3 mm data are larger by a factor of \about2 
than those from 0.87 mm data partly because of the larger beam size (and dilution) of the latter data.
Other possible reasons for any discrepancy between the parameters derived from 0.87 mm and 1.3 mm data 
are deviation of dust temperature and $\beta$ from the adopted values,
different \uv\ coverages, 
small contamination to the continuum from non-thermal or line emission, 
and our flux calibration errors.

The overall uncertainties of our estimates for the peak surface densities and masses are about
a factor of 5, considering the uncertainties of the opacity coefficient, gas-to-dust ratio, 
dust temperature, and calibration of our observations.
The largest source of error is in the dust opacity coefficient. 
\citet{Pollack94} estimated its very conservative error to be 
a factor of 4 for a variety of grain properties. 
The gas-to-dust ratio that we adopted is from the ratio in the solar neighborhood \citep[$\approx$150;][]{Sodroski97}
corrected for the metallicity at the center of NGC 253, 
which is 0.1--0.2 dex higher than in our vicinity \citep{Pilyugin04,Pilyugin06}.
The dust temperature in the calculation above is from the peak brightness 
temperature of the \twelveCO\ lines at the continuum sources.
It is also consistent, within a factor of 2, with the dust temperatures inferred for the nuclear 
region by fitting the spectral energy distribution (SED) in the far-infrared \citep{Radovich01,Melo02,Weiss08}.
We did not subtract non-dust emission from the continuum or correct the continuum flux density
for missing flux because the two corrections are less than a few 10\% each 
and are in opposite directions.

\subsubsection{Association with Known Sources}
\label{s.continuum.assoc}
The continuum peaks at $\lambda \sim 1$ mm are spatially well correlated with those in centimeter
wavelengths, as illustrated in Fig. \ref{f.cont} (c). 
Four of the five peaks of dust emission have one or more counterparts of centimeter compact
sources, in particular the bright centimeter peaks in the nuclear region customarily called TH$n$ after \citet{Turner85}. 
The brightest 1.3 mm peak, No. 3 in Fig. \ref{f.cont} (c), 
is associated with the brightest centimeter source TH2, which is often presumed
to be the radio nucleus and is within 2\arcsec\ of the stellar kinematical center \citep{Muller-Sanchez10}.
This millimeter source is elongated along the position angle of \about60\arcdeg\ in 
agreement with the elongated distribution of compact radio sources observed 
by \citet{Turner85} and \citet{Ulvestad97} in the immediate vicinity of TH2.
Since the radio sources comprise supernova remnants and \ion{H}{2} regions, as well as the possibly
active nucleus of low luminosity, the coincidence of centimeter and dust emission peaks suggests that
the 1 mm peaks are dust (and gas) condensations heated by those sources (i.e., mostly massive stars).

In addition to the centimeter-wave continuum sources, 
two \water\ masers in the galactic center \citep{Henkel04,Hofner06} are spatially associated with
two of our continuum peaks.
Also, the most luminous infrared source in the galactic center, attributed to a super star cluster
\citep{Keto99,Kornei09}, is at the millimeter emission peak No. 4 according to its spatial association
with the radio source TH7 \citep{Kalas94,Dudley99,Fernandez-Ontiveros09}.
Our 1.3 and 0.87 mm continuum sources are also associated with peaks in the molecular line emission 
described below.

\subsection{Line Emission Properties} \label{s.line}
Molecular line maps in Figs. \ref{f.integ} and \ref{f.maxTb} show 
peaks corresponding to the five dust continuum peaks,
less prominent gas clumps, and extended line emission.
The observed parameters of the line emission peaks are listed in Table \ref{t.line_peaks}.

\subsubsection{CO Excitation and Gas Temperature} \label{s.line.excitation_Tb}
We first note that the peak integrated intensities as well as the peak brightness 
temperatures are approximately the same in the two transitions of \twelveCO\
at the five most prominent peaks of dust and gas emission.
This suggests that the bulk of the \twelveCO\ emission from the line-emission peaks, 
or molecular cloud complexes, is optically thick and is nearly thermalized at least 
to the $J=3$ level.
Under this condition of (near) local thermal equilibrium (LTE), 
the brightness temperature of an optically thick line gives the gas temperature 
at the cloud photosphere multiplied by the beam filling factor of  the emission.

\begin{figure}
\epsscale{0.57}
\plotone{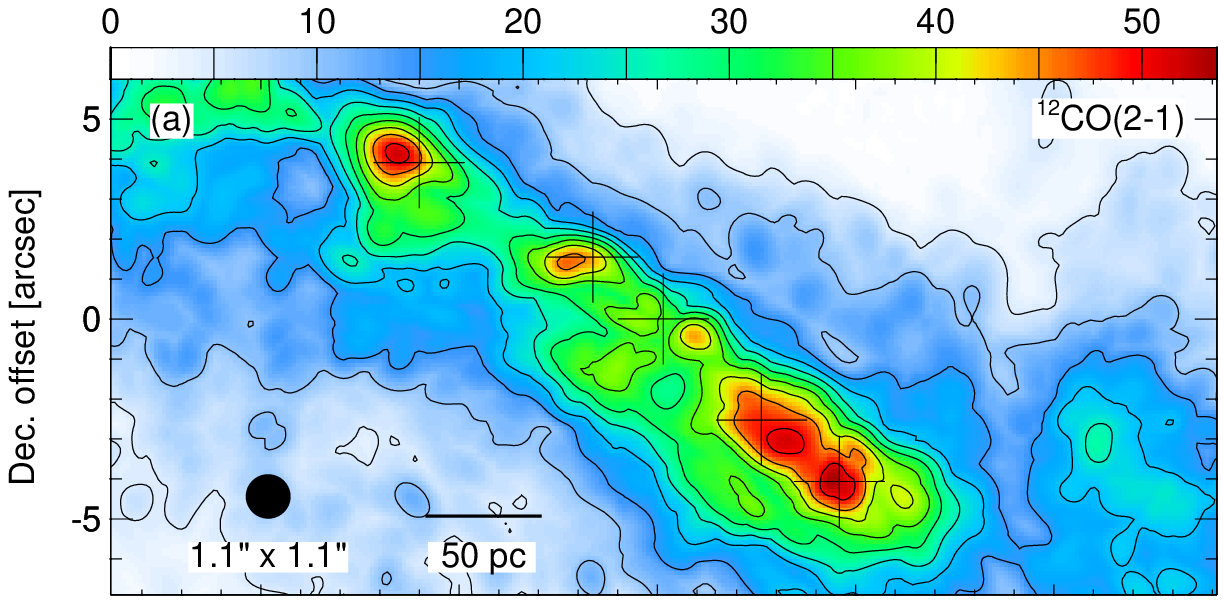} \\ 
\plotone{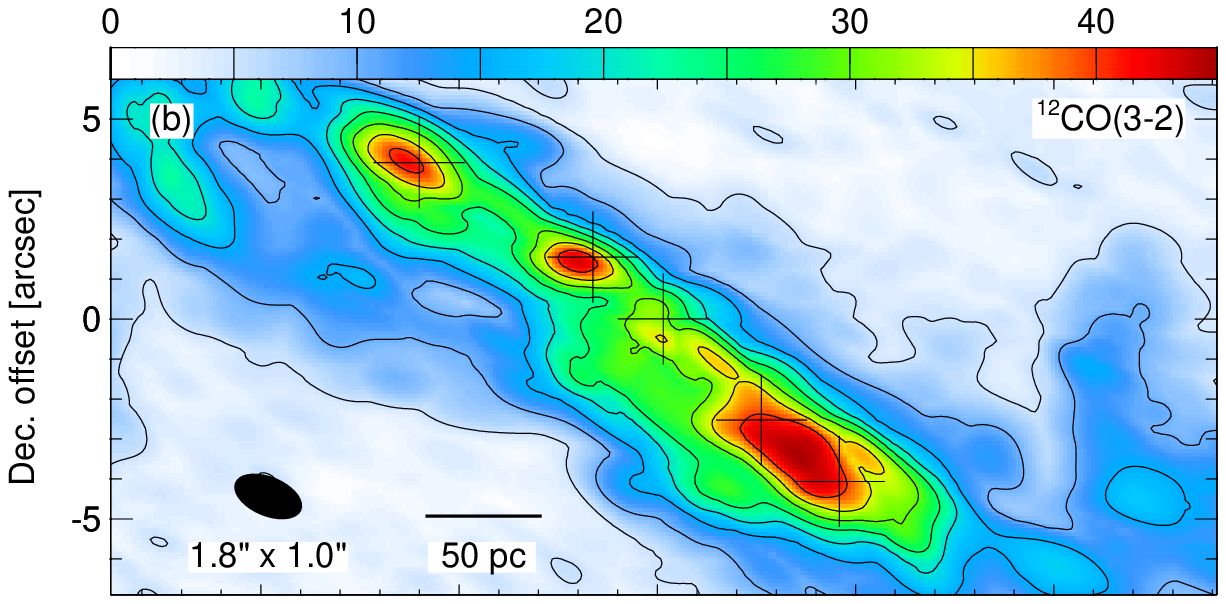} \\ 
\plotone{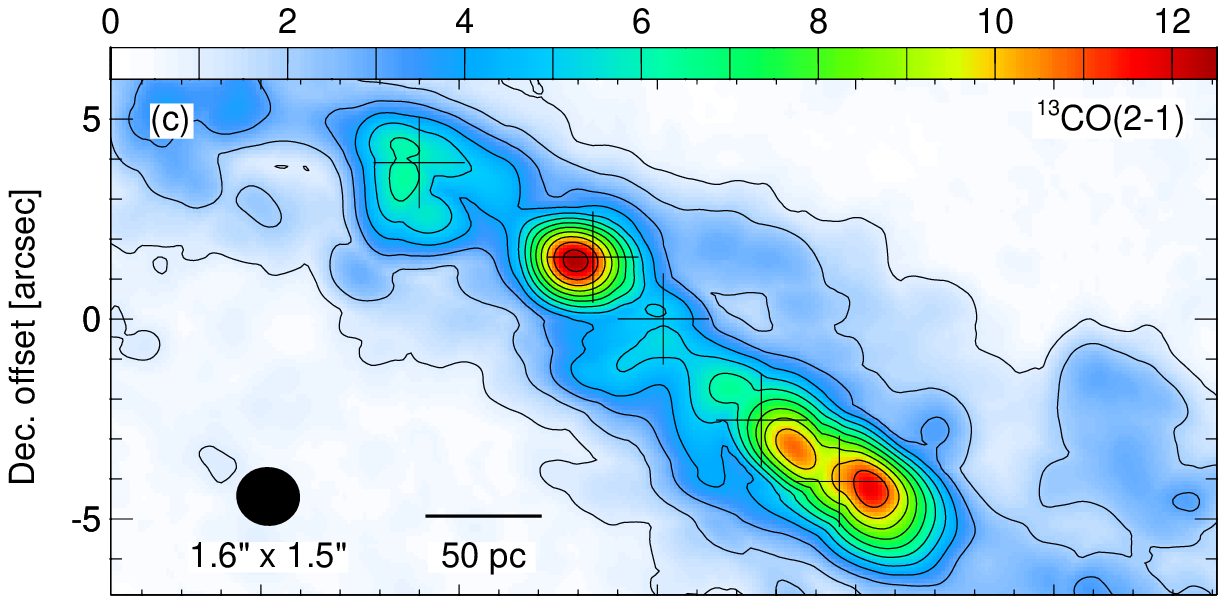} \\ 
\plotone{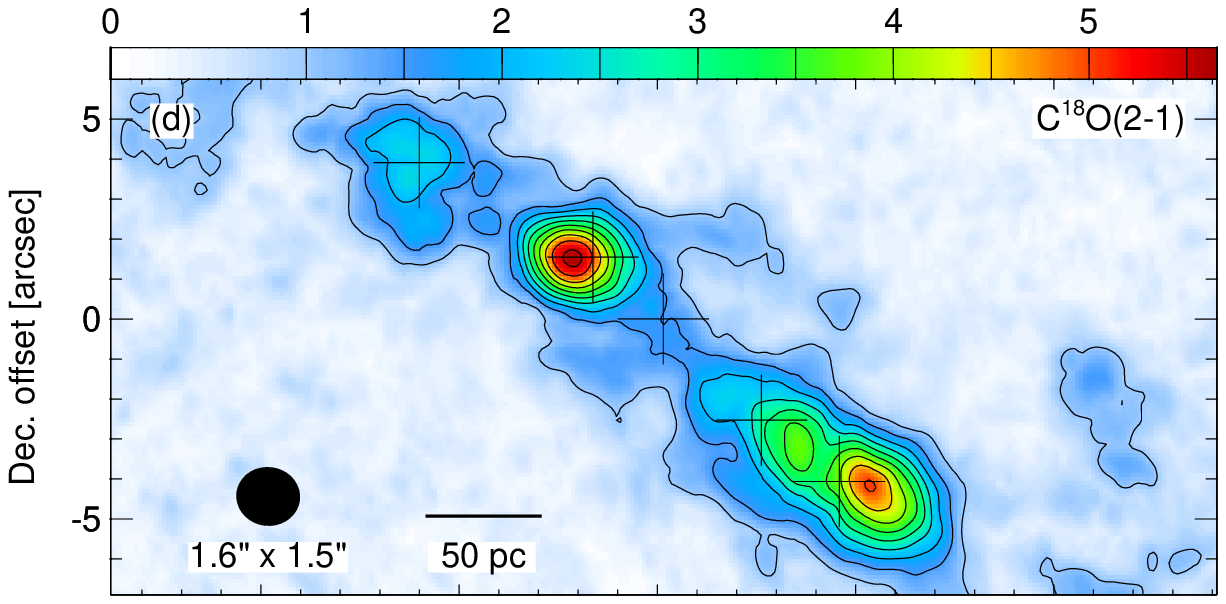} \\ 
\plotone{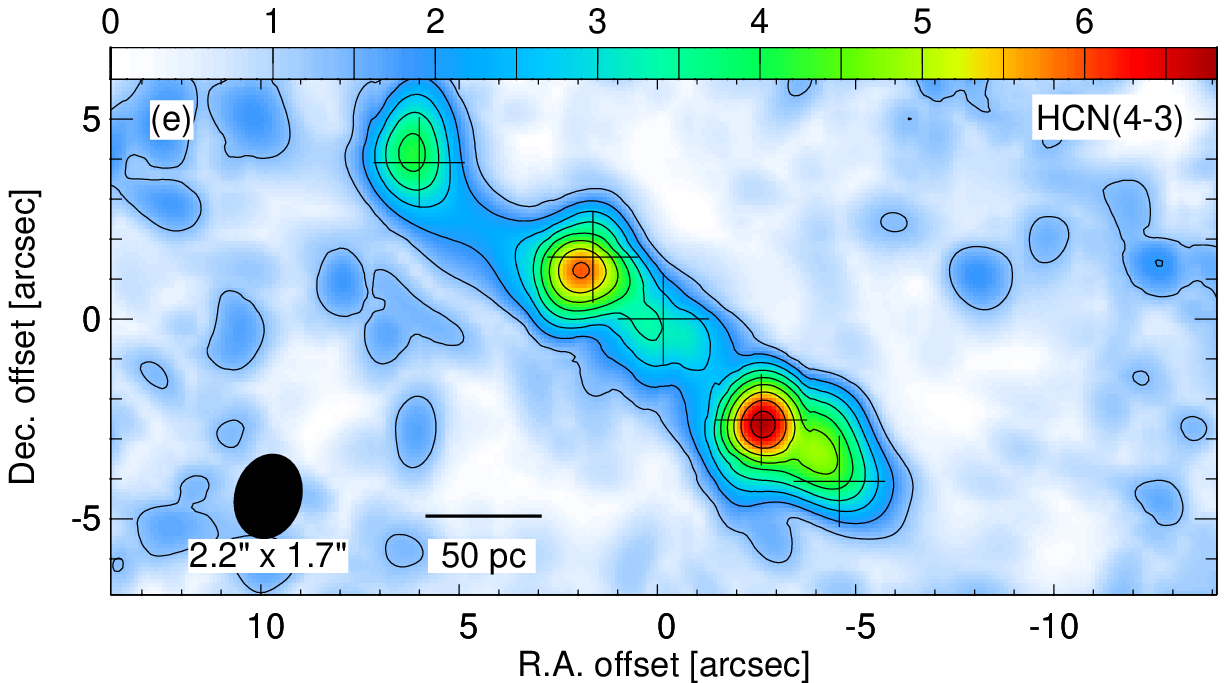} \\ 
\epsscale{1.0}
\caption{Peak brightness temperature maps of molecular lines in the center of NGC 253. 
The Rayleigh-Jeans brightness temperatures of the lines (in excess of any continuum) are in units of K. 
The five plus signs are at the positions of the 1.3 mm continuum peaks.
\label{f.maxTb} }
\end{figure}

The peak brightness temperature (\Tb) of the \twelveCO\ lines, shown in Fig. \ref{f.maxTb}, 
has maxima of about 50 K  at or in the vicinity of the integrated intensity peaks.
These beam-averaged brightness temperatures set a lower limit to the beam-averaged temperature 
of molecular gas at each location.
Our data therefore suggest that molecular gas at and around four of the five 1 mm peaks
has a minimum beam-averaged temperature of 50 K.
This temperature is not much higher than our previous lower limit of 40 K based on the \about3\asec\
resolution observations of \citest{Sakamoto06},
suggesting that the extent of individual CO-emitting cloud complexes is about  1\asec--3\asec\ (20--50 pc).
An exception to the \about50 kelvin \Tb\ among the five peaks is the galactic nucleus, 
i.e., peak No. 3, where peak \Tb\ is about 30 K in both \twelveCO\ lines.
This is quite likely due mainly to the fast gas motion around, and large velocity gradient across,  
the dynamical center
since gas has a small filling factor and a large beam dilution in such a location.
It is also notable that the region with peak \Tb\ higher than 30 K, which is about 3 times higher than
the peak brightness temperatures of giant molecular clouds in the disk of our Galaxy \citep{Scoville87}, 
has an extent of 300 pc by 50 pc (17\asec\ $\times$ 3\asec) in Fig. \ref{f.maxTb} (a).
The high temperature in the CMZ of NGC 253 is therefore not limited to the most prominent emission peaks.
This notion is strengthened when we consider that the beam filling factor of gas is likely lower 
at locations further away from the peaks of integrated intensities.

\subsubsection{Gas Column Density, Mass, and CO Opacity}  \label{s.line.column_mass_opacity}
The peak column density of gas at the molecular cloud complexes is estimated to be 
\about$10^{4}$ \Msol\ \persquarepc\ from the line emission using multiple methods that we describe below.
This is in agreement with the estimate from dust continuum in \S \ref{s.continuum.mass}.
The gas mass of each molecular complex is therefore on the order of $10^{7}$ \Msol\ 
for the clumps identified in our $\approx$20 pc resolution data.

Our three methods to estimate gas surface density from the line data are the following.
Firstly, the surface density of molecular gas is
$\Smol=1\times10^{4}$ \Msol\ \persquarepc\ 
for the \twelveCO\ integrated intensity $I_{\rm ^{12}CO} = 5\times 10^{3}$ K \kms, 
a typical peak value at the five peaks in both CO(3--2) and (2--1), 
for the  \twelveCO--\HH\ conversion factor of 
$\Xco = 1 \times 10^{20} $ \persquarecm\ (K \kms)$^{-1}$
and the factor of 1.4 correction for heavy elements.
Here we did not make a distinction between the \twelveCO\ lines that we observed and
\twelveCO(1--0) for which \Xco\ is most often estimated, since $I_{\rm ^{12}CO(1-0)}$ should be 
approximately equal to $I_{\rm ^{12}CO(2-1)}$ and $I_{\rm ^{12}CO(3-2)}$ for the excitation condition
that we suggested most likely.
The \Xco\ factor is usually estimated to be in the range of (0.3--3)$\times 10^{20}$ \unitofX\
for large metal rich galaxies \citep[e.g.,][and references therein]{Arimoto96}.
For the center of NGC 253, \Xco\ has been estimated to be
4, 1.7, 1.3, and 0.3 in units of $10^{20}$ \unitofX\ by
\citet{Israel95}, \citet{Weiss08}, \citet{Martin10}, and \citet{Mauersberger96}, respectively.
Secondly, 
the peak gas column densities at the molecular clumps are estimated to be
$\Smol \approx 10^{4.0}$ \Msol\  \persquarepc\ from $I_{\rm ^{13}CO(2-1)} \approx 800$ K \kms\ and
$\Smol \approx 10^{4.2}$ \Msol\  \persquarepc\ from $I_{\rm C^{18}O(2-1)} \approx 300$ K \kms\
using 
the conversion relations from line flux to gas column density
$\Smol = 10^{1.1} I_{\rm ^{13}CO(2-1)}$ and
$\Smol = 10^{1.7} I_{\rm C^{18}O(2-1)}$, 
where \Smol\ is in \Msol\  \persquarepc\ and the line flux K \kms.
Assumptions used here are that
the lines are optically thin and from gas in LTE at 50 K
and that the molecular abundances are 
[\twelveCO/\HH]=$10^{-4}$, [\twelveC/\thirteenC]$\approx$80, and [\sixteenO/\eighteenO]$\approx$300;
the latter two are from the lower limits in the center of NGC 253 obtained by \citet{Martin10}. 
Thirdly, the peak gas column density for $I_{\rm C^{18}O(2-1)} \approx 300$ K \kms\ is
$\Smol \approx 10^{3.9}$ \Msol\  \persquarepc\ if we use the
 relation $\Smol = 10^{1.4} I_{\rm C^{18}O(2-1)}$
 \citep{Stutzki90,Mauersberger92,Wild92}.
 Here the assumptions are that the \CeighteenO\ line is optically thin, 
 that the gas is warm and moderately dense
(i.e., $T_{\rm kin} > 40$ K and $10^3 < n_{\rm H_{2}} < 10^{5.5}$ \percubiccm),
and that the molecular abundances are the same as the one assumed above.
In this third method, LTE is not assumed. 
Indeed, the factor of 2 lower estimate than the LTE value is because this method takes into account
the difficulty of populating levels higher than $J=2$.

The optical depths of CO lines are
15, 0.17, 0.05, and 25 for
\twelveCO(2--1), \thirteenCO(2--1), \CeighteenO(2--1), and \twelveCO(3--2), respectively,
for a 50 K gas in LTE with a line width of 100 \kms\ and
the surface density of
$\Smol=1\times10^{4}$ \Msol\ \persquarepc\ 
(i.e., \HH\ column density $N_{\rm H_2} = 5\times 10^{23}$ \persquarecm).
These calculated opacities should be representative of the line opacities of the cloud complexes
because the parameters used are the ones we estimated or adopted for the cloud complexes.
The above opacities for \thirteenCO\ and \CeighteenO\ are indeed comparable to the
observationally estimated mean opacities of these lines in Table \ref{t.line_peaks}.
Unless \thirteenC\ and \eighteenO\ are significantly overabundant than we assumed,  the \twelveCO\ lines 
that we observed in the molecular cloud complexes are optically thick.
The  \thirteenCO(2--1) and \CeighteenO(2--1) lines are optically thin on average
although there may well be small regions of high column density where these lines are opaque,
because the gas distribution in each molecular complex is almost certainly non-uniform.

\subsubsection{Line Profiles and Self-absorption of \twelveCO} \label{s.line.profile_selfabsorption}
\begin{figure}
\epsscale{0.9}
\plotone{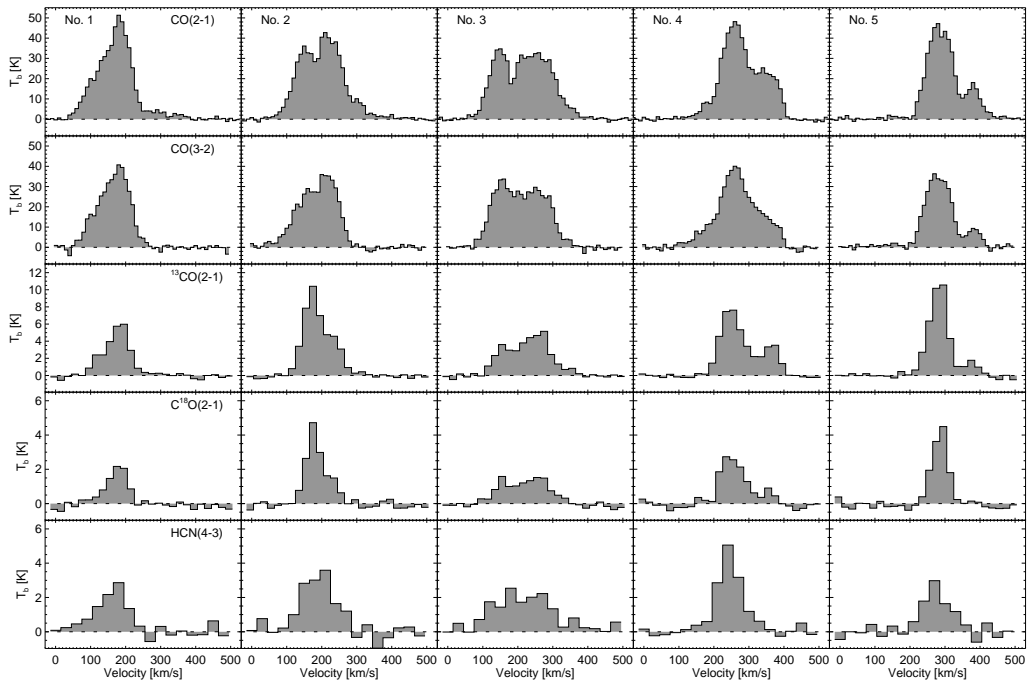}  
\epsscale{1.0}
\caption{Line profiles at the five most prominent molecular cloud complexes in the galactic center. 
Each spectrum is from the cloud position in Table \ref{t.line_peaks} and is corrected for the primary beam attenuation.
The variation of line shapes between different molecular lines, 
seen in the molecular complexes No. 2, 3, 4, and 5,
suggests self-absorption of \twelveCO\ or multiple velocity components.
\label{f.ispec5in1} }
\end{figure}

The line profiles of the five most prominent molecular complexes are shown in Fig. \ref{f.ispec5in1}.
The line widths are around 100--200 \kms.
The line shapes differ from one complex to another and sometimes from one line to another
for the same molecular complex.
The broad line-widths and the variation of line shape among the cloud complexes
must be partly due to the galactic rotation and the large-scale non-circular motion
in the barred galaxy as seen in the mean velocity map (Fig. \ref{f.12co21.mom1}) rather than due
solely to velocity dispersion of molecular gas.

Regarding the variation of line profile between different lines, we note the cases of the molecular
complexes No. 2 and 3. 
In the former the \twelveCO(2--1) and (3--2) profiles have a less sharp peak than the \thirteenCO\ and
\CeighteenO\ profiles. 
A dip is in \twelveCO\ profiles (or at least in the $J$=2--1 profile) at the peak velocity of
\thirteenCO\ and \CeighteenO. 
In the latter \twelveCO(2--1) has a sharp dip that is not seen in other lines. 
We suggest these to be self-absorption of \twelveCO, since, as we evaluated above, 
\twelveCO\ is optically thick and \thirteenCO\ and \CeighteenO\ optically thin. 
Since the continuum emission has peak brightness temperatures of less than 2 K at these locations,
any absorption much deeper than 2 K must be mostly toward brighter CO emission rather than continuum.
The self-absorption then means that there is colder gas in front of warmer gas along our line of sight.
Each absorber can be either inside or outside of the cloud complex
but is probably within the central molecular zone considering the
78\arcdeg\ inclination of NGC 253.
The waker (or lack of)  absorption in CO(3--2) can be partly because the foreground gas is
not warm and dense enough to have a significant population at $J$=2.

An important implication of the self-absorption is that the peak brightness temperature, 
column density, and mass of the molecular cloud complexes will be underestimated using \twelveCO\
for the absorbed gas. 
For example, the molecular cloud complex associated with the
millimeter continuum peak No. 2 may have the largest column density among the five complexes.
It is the brightest among the five in the optically thinnest \CeighteenO\ line, even though it 
has comparable or less integrated intensities in \twelveCO.

In addition to the self-absorption, there are probable cases of 
cloud overlapping along our line of sight.
For example, the molecular complexes No. 4 and 5 have double-peaked line profiles. 
These are likely due to multiple velocity components because the double peaks 
are visible even in \thirteenCO\ and, in the case of the complex No. 4, \CeighteenO.
The CMZ must be a thin disk-like volume even though
the most prominent dust and gas features are on an one-dimensional string on the sky from our 
near edge-on viewpoint.
Thus there must be substantial overlap of molecular clouds along our lines of sight.

\subsubsection{Gas Density and Line Ratios} \label{s.line.ratio_density}
The {\em volume-averaged} gas density is estimated to be $\nhtwo \approx 10^{3.6}$ \percubiccm\
for a 20 pc diameter region around the molecular emission peaks. 
This is from the peak gas column density of $10^4$ \Msol\  \persquarepc\ 
and an assumption that half of the column density is from within the 20 pc.
This density is almost as high as the critical density for CO excitation for the $J=2$ level, 
$10^{4}$ \percubiccm\ for 50 K.
Since \twelveCO\ excitation has additional help from photon trapping, 
it is almost certain that \twelveCO\ is thermalized at $J \approx 2$ 
as we inferred from the CO(2--1)/(3--2) ratio. 
Our volume-averaged gas density is comparable to the density of $\nhtwo \sim 10^{3.9}$ \percubiccm\
that \citet{Gusten06} estimated for the CO-emitting molecular gas in the central 15\arcsec\ (250 pc) 
from LVG analysis of multi-transition CO data.
This agreement between macroscopic and microscopic gas densities
suggests that the \twelveCO(2--1) emitting gas has a large volume filling factor 
in the molecular cloud complexes.
When the molecular gas containing \twelveCO\ almost fills the volume of each molecular emission peak,
the effect of beam-dilution must be small, unless there is a large systematic velocity gradient 
at the emission peak. 
Thus the peak brightness temperatures of $\approx$50 K that we measured in \S \ref{s.line.excitation_Tb} 
must be a good estimate of the mean gas temperature at the \twelveCO\ photosphere 
(with the likely exception of the molecular peak No. 3 at the galactic dynamical center.)

\begin{figure}
\epsscale{0.57}
\plotone{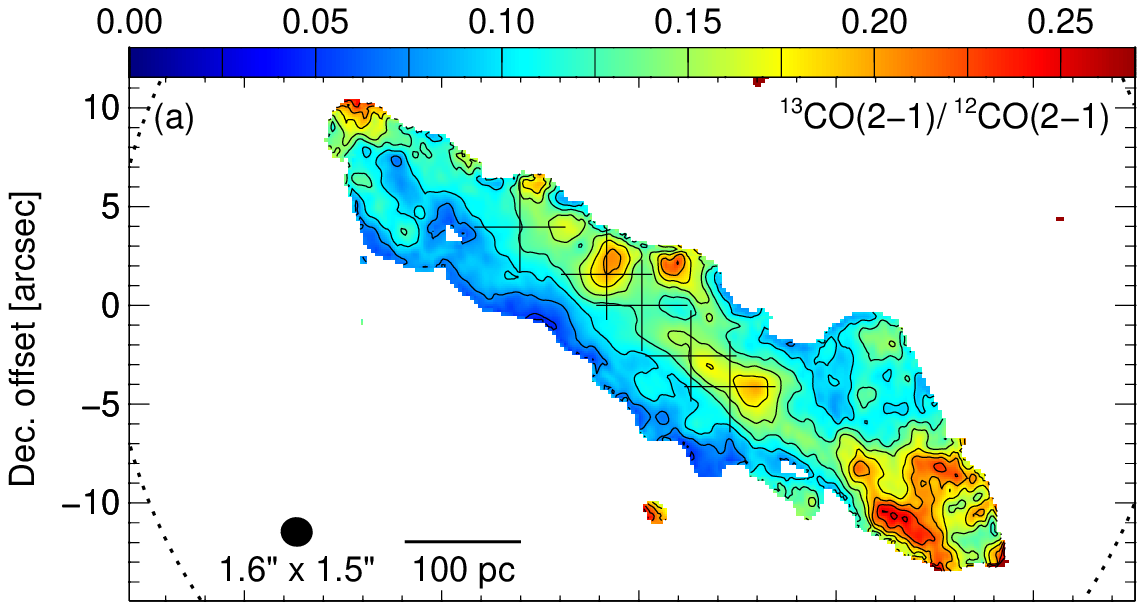} \\ 
\plotone{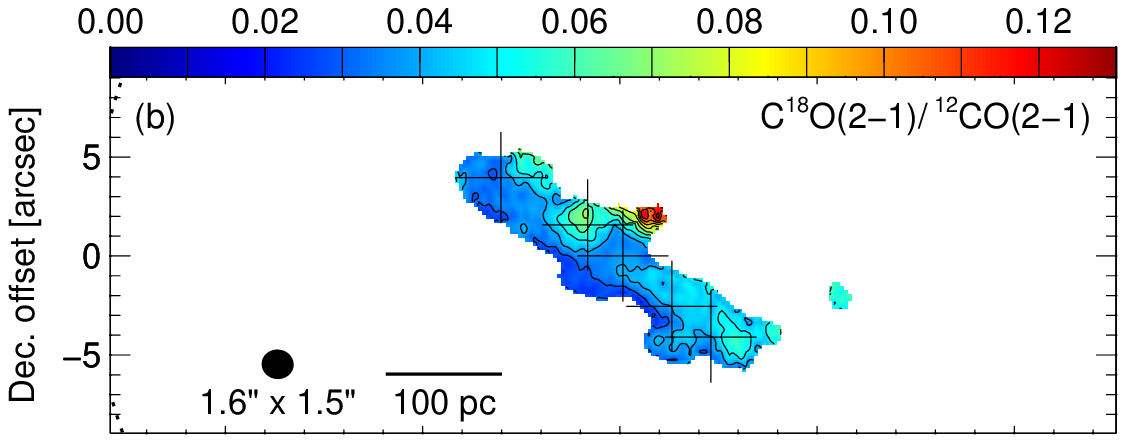} \\ 
\plotone{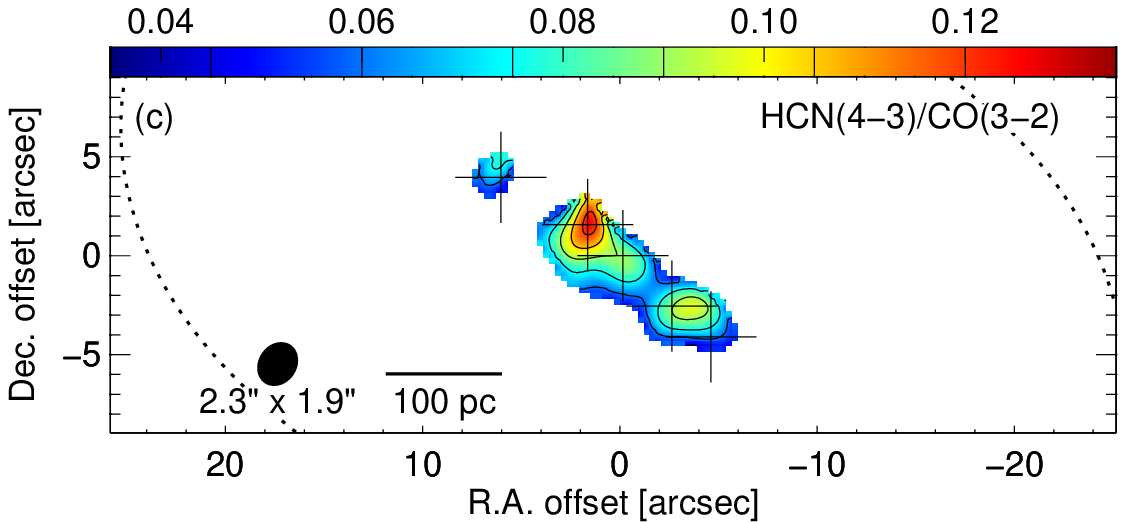} \\ 
\epsscale{1.0}
\caption{Ratio of line integrated intensities. 
(a) \thirteenCO(2--1) to \twelveCO(2--1).
(b) \CeighteenO(2--1) to \twelveCO(2--1).
(c) HCN(4--3) to CO(3--2).
For each ratio, the two integrated intensity maps were convolved to the same resolution
and expressed in units of K \kms\ before calculating the ratio at locations where
both lines are detected above 5 $\sigma$.
Plus signs show the locations of the five 1.3 mm continuum peaks.
\label{f.ratio} }
\end{figure}

The intensity ratios of molecular lines are not uniform in the CMZ, 
suggesting spatial variation of gas properties in the region. 
Such variations are most obvious in the HCN-to-CO ratio and in the ratio between \twelveCO\ and
other less abundant molecules (Fig. \ref{f.ratio}). 
The HCN-to-CO intensity ratio is higher at the chain of molecular complexes, 
0.06--0.08 at the \twelveCO(3--2) peaks, than in other parts of the CMZ, where
the mean ratio is 0.04 in the central 20\asec.
To put it differently, 83\% of HCN(4--3) flux detected in the central 20\asec\ (340 pc, FWHM) is from
3\asec\ (50 pc) diameter regions at the five continuum peaks
while the fraction is only 30--44\% for the four CO lines.
The HCN emission is clearly more concentrated in the five molecular complexes.
The ratio map Fig. \ref{f.ratio}(c) shows that the ratio has a peak value of about 0.12 near the
millimeter continuum peak No. 2.
HCN emission generally traces dense molecular gas because of the large dipole moment of the molecule
and hence its large critical density.
Multi-transition analyses of HCN in the CMZ of NGC 253 suggested 
that HCN molecules are subthermally excited 
and that the most plausible density of the HCN-emitting gas  is $n_{{\rm H}_2} \sim 10^{5}$ \percubiccm\
\citep{Knudsen07,Paglione97,Jackson95}.
Thus the concentration of HCN(4--3) emission in the molecular complexes strongly 
implies a large concentration of dense gas there, possibly in the form of many dense gas cores 
embedded in giant molecular cloud complexes.
HCN enhancement at the largest cloud complexes may have some additional causes
such as  infrared pumping of HCN \citep{Carroll81},
excitation by electron collisions \citep{Dickinson77}, 
and spatial variation of chemical abundance.
Another spatial variation is seen between \twelveCO\ and less abundant molecules.
Specifically, the galactic nucleus is the strongest peak in the integrated
intensity maps of \twelveCO\ (2--1) and (3--2) while the two adjacent molecular complexes
are stronger in less abundant \thirteenCO, \CeighteenO, and HCN (see Figs. \ref{f.integ} and \ref{f.ratio}). 
It is interesting to note that gas around the continuum peak No. 2 again has a peak  in the CO isotopologue ratios.
The off-nuclear peaks also dominated, and often produced double-peak appearance, 
in previous high-resolution ($\lesssim$ 5\asec) observations of non-CO molecules
(HCO$^{+}$ and HCN, \citet{Knudsen07}; 
H$_2$S, \citet{Minh07}; 
SiO and H$^{13}$CO$^{+}$, \citet{Garcia00};
CS, \citet{Peng96}).
This difference between the nuclear and off-nuclear gas peaks may be 
because the gas temperature is higher at the nucleus 
as we will see in \S\ref{s.line.line_cont.comparison} from the line-to-continuum ratio.

\subsubsection{Other Line Features}
\begin{figure}
\epsscale{0.57}
\plotone{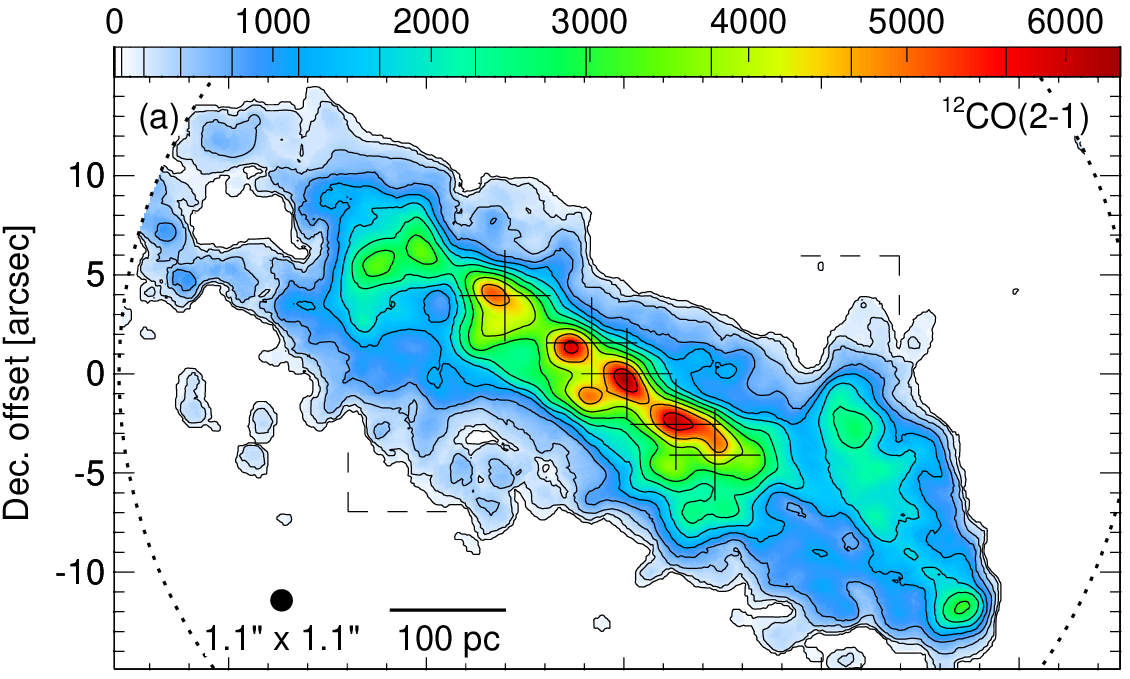} \\ 
\plotone{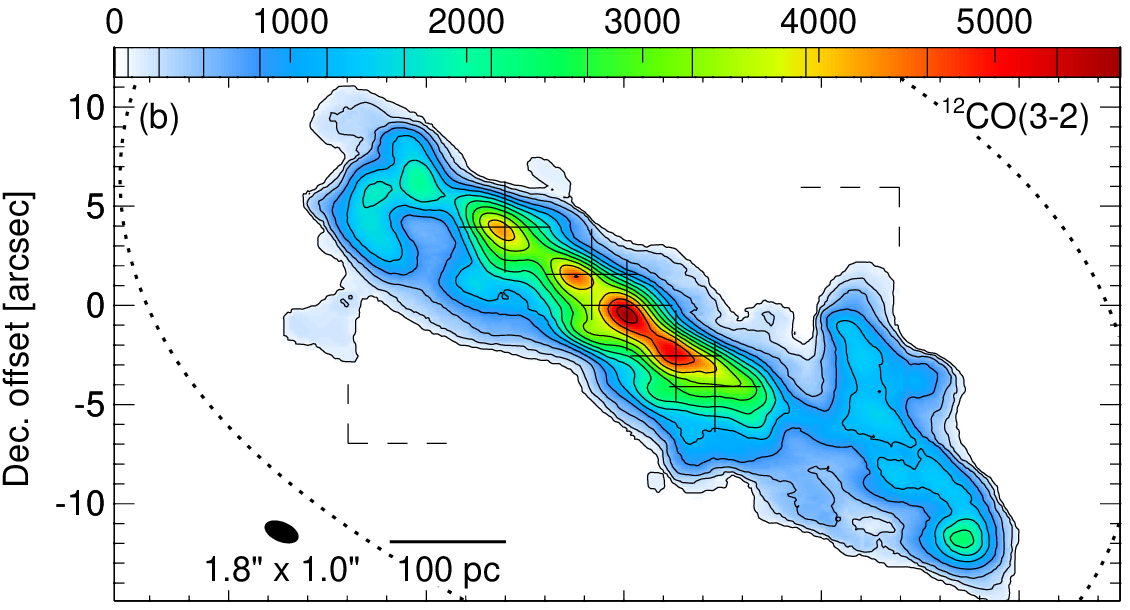} \\ 
\plotone{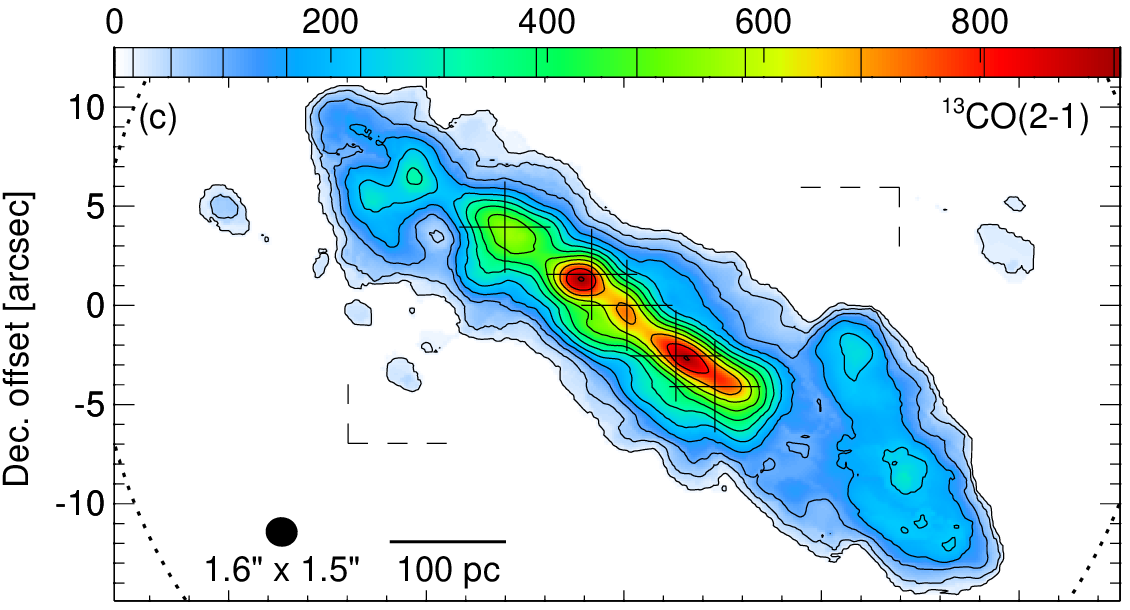} \\ 
\plotone{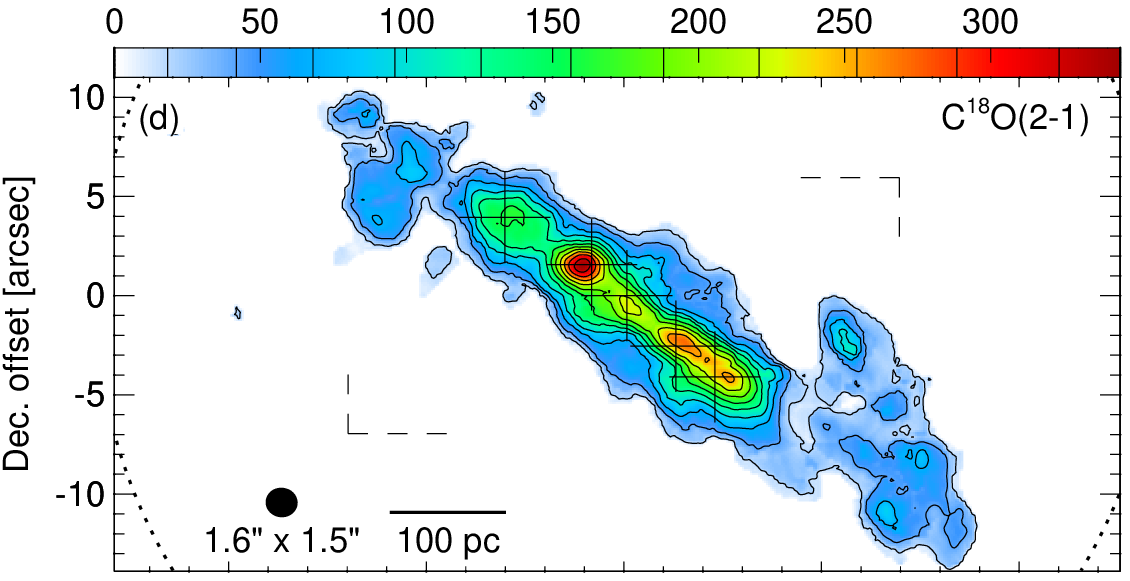} \\ 
\plotone{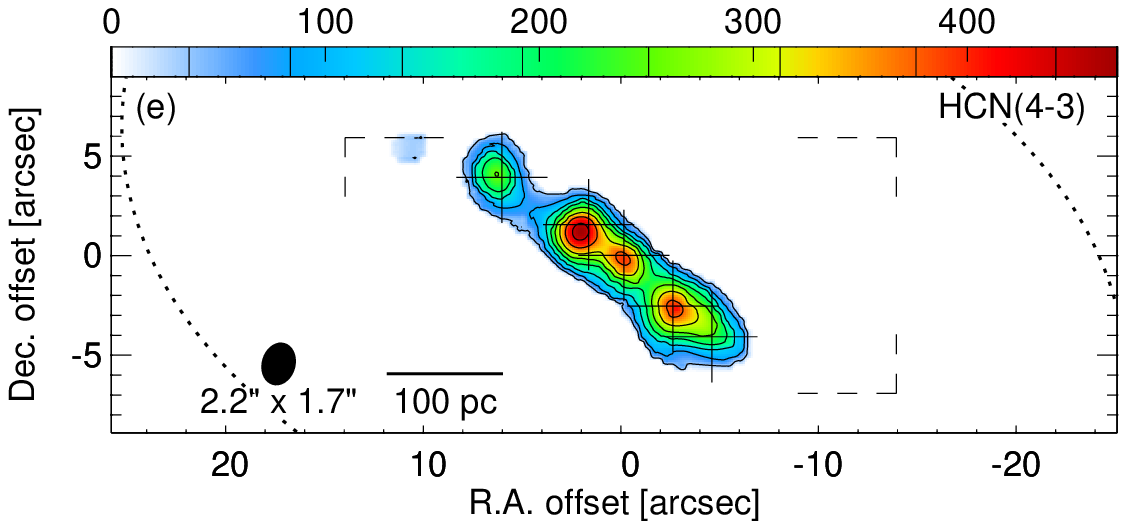} \\ 
\epsscale{1.0}
\caption{\small
Molecular lines in the center of NGC 253.
The line name and the synthesized beam (FWHM) are shown in the
top-right and bottom-left corners, respectively.
The five plus signs are at the positions of the 1.3 mm continuum peaks.
The dotted lines are the half-power contours of the (mosaicked) primary beams, for which
the maps are corrected.
The dashed lines mark the area shown in Figs. \ref{f.cont} and \ref{f.maxTb}.
The unit of integrated intensity is K \kms.
The $n$-th contour is at $46\times n^2$, $78\times n^{1.7}$,
$17\times n^{1.6}$, $18\times n^{1.2}$, and $36\times n^{1.2}$,
respectively, from (a) to (e).
\label{f.integ} }
\end{figure}

Our line data also show features that were less clearly seen in previous works.
First, the asymmetry of \twelveCO\ emission about the major axis, pointed out in \citest{Sakamoto06},
is more evident in the new data. 
For example, in the \twelveCO(2--1) map in Fig. \ref{f.integ}, 
the contours at around 2000 K \kms\ are clearly asymmetric about the major axis
that runs approximately through the continuum peaks marked with crosses. 
The north-western side of the nucleus, which is the side near to us, shows sharper outward decline 
in \twelveCO\ intensity than the south-eastern side, which is the far side of the galaxy. 
As inferred in \citest{Sakamoto06}, this is most likely due to high opacity and area filling factor
of \twelveCO-emitting gas because the asymmetry is seen only in \twelveCO.
The asymmetry in optically thick lines can be caused by the vertical thickness of the CMZ
and the plausible temperature gradient of molecular gas (i.e., lower temperature at the larger radii).
It is because, for a disk with thickness $2h$ and inclination $i$,
the center of the disk surface is offset by $h \sin i$ to the far side 
from the center of the disk volume when projected onto the sky.
The former center is hottest in the photosphere of an optically thick line while the latter is the 
true center of the disk and the hottest point for an optically thin emission.
The offset can be enhanced if the molecular disk has the shape of a biconcave lens, which is observed 
in the Galactic CMZ \citep[e.g.,][]{Bally10}. 
The starburst wind from the CMZ of NGC 253 \citep[e.g.,][]{Lehnert96} may be a reason for such a gas distribution. 
Similar near-side/far-side asymmetry in molecular lines has been reported in the centers of 
\objectname{NGC 1068} \citep{Baker98}, 
\objectname{IC 342} \citep{Meier01}, 
\objectname{NGC 1365} \citep{Sakamoto07}, and 
\objectname{NGC 4826} \citep{Boone11};  
the last paper has more of possible cases and a model of the opacity effect.

Second, one of the bubbles noted in \citest{Sakamoto06}, called SB2, is seen in our \twelveCO(2--1)
map at $(\Delta \alpha, \Delta \delta) \sim (20\asec, 8\asec)$ with respect to the radio source TH2
(see Fig. \ref{f.integ}a),
while SB1 at around $(-9\asec, -7\asec)$ is not evident in the integrated intensity map. 
The gas distribution at the position of SB1 is indeed disrupted in both channel maps and  
the integrated map,  but, unlike SB2, the feature does not appear to be semi-circular in our new data.
The nature of SB1 thus remains uncertain; it can be an expanding bubble in a complicated shape
or other gas kinematical feature that looked like a bubble in low resolution.
The third possible bubble inferred in \citest{Sakamoto06} from position-velocity diagrams is identified with a
depression at $(9\farcs5, 3\farcs5)$ in the integrated intensity map (e.g., Fig. \ref{f.integ}c).

Third, the velocity map in Fig. \ref{f.12co21.mom1} confirms the non-circular motion of molecular gas 
in the CMZ since the isovelocity contours run almost {\it parallel} to the major axis of the galaxy 
at P.A. \about 50\arcdeg. 
Our data also reveal systematic deviation of the velocity field from the rest of the CMZ 
in the central few arcsec around the nucleus (or TH2).
Velocity contours are kinked in that region 
in such a way that the kinematical major axis approximately agrees with the galaxy major axis.
This gaseous velocity field, in particular the strong noncircular motion in the CMZ, is in
sharp contrast to the circular velocity field of stars observed in the same region by 
\citet{Muller-Sanchez10}.  
Because of this large non-circular motion (i.e., lack of the usual spider pattern in velocity contours),
the current data do not distinguish whether the dynamical center of the galaxy is at the radio
source TH2 or \about0\farcs7 from it as estimated by \citet{Muller-Sanchez10}.

\subsection{Comparison of Line and Continuum} 
\label{s.line.line_cont.comparison}
The molecular cloud complexes as well as their brightness temperature peaks are spatially associated 
with the \about1 mm continuum peaks and also with the centimeter-wave and infrared
sources in and around them (\S \ref{s.continuum.assoc}).
However, there are sometimes slight offsets of $\lesssim1$\asec\  between line and continuum peaks.
This can be seen in the line emission maps in Figs.  \ref{f.integ} and \ref{f.maxTb} where the millimeter
continuum peaks in Fig. \ref{f.cont} and Table \ref{t.cont_peaks} are plotted as crosses.
For example, the 1 mm peak No. 2 is offset about 0\farcs5--1\asec\ west or northwest from
the nearest peak in our line emission maps. 
This is seen even in optically thin \CeighteenO\ as a 0\farcs5 offset in both Fig.  \ref{f.integ}d and \ref{f.maxTb}d
and is therefore unlikely to be due to line opacities.
The line-continuum offset is less significant for the peak No. 3 at the nucleus.
A possible reason for the line-continuum offset is that the dust emission follows
the heating from massive stars more truly than the molecular line emission does
because dust can absorb UV radiation and be heated while molecules are dissociated
by the same radiation.
If this is the case the offsets imply that active star formation occurred at the periphery 
of the molecular complexes or that the stars formed in groups are drifting
away from the molecular complex.

Regarding the intensity comparison between line and continuum emission,
we have one note each for optically thick and thin lines.
For optically thick \twelveCO, the line-to-continuum ratio, or line equivalent width, is roughly uniform
among the peaks No. 2 through 5 and is smaller than the mean value of the CMZ. 
(The peak No. 1 is too weak for this analysis.)
The \twelveCO(3--2) equivalent  widths are (3--4)$\times 10^3$ \kms\ 
while the mean value in the central 15\arcsec\ is $(15\pm3) \times 10^3$ \kms\   
\cite[][and references therein]{Sakamoto08}.
The smaller equivalent widths in these peaks are consistent with their high column densities. 
The simple LTE model in \citet{Sakamoto08} suggests column densities of $N_{\rm H_2}/\Delta V \sim 10^{22}$
\persquarecm \perkms, which agrees within a factor of a few with our estimate 
in \S \ref{s.line.column_mass_opacity}.

For optically thin CO lines, 
the line-to-continuum ratio shows striking variation among the
five peaks in contrast to the case of opaque \twelveCO.
The peak No. 3 at the nucleus is the brightest in the continuum but 
weaker than other peaks in \thirteenCO\ and \CeighteenO, as
already noted in our previous observations \citesp{Sakamoto06}.
One of the possible reasons discussed in \citest{Sakamoto06} was 
significant contribution from synchrotron or free-free emission to the continuum.
This is not supported by the newly measured spectral index of the continuum at this peak, about 3,
that is comparable to those at other peaks. 
A remaining plausible explanation for the stronger continuum at the nucleus is a higher temperature than in
the other peaks. 
A higher temperature reduces the equivalent widths of optically thin lines but
does not change the equivalent width of a thick line in the LTE model 
\citep{Sakamoto08}. 
The higher temperature at the nucleus is thus consistent with our observations of CO isotopologues.
Presence of molecular gas at $\gtrsim$100 K in the central region of NGC 253 has been inferred by
observations of \HCthreeN\ \citep{Mauersberger90}, \ammonia\ \citep{Mauersberger03}, high-$J$ CO lines \citep{Bayet04}, and the combination of CO isotopologues and HCN \citep{Matsushita10}.
A large fraction of such hot molecular gas may be at the nuclear peak (i.e., No. 3) causing the low
equivalent width of optically-thin CO lines there.
Another remaining model is the selective photodissociation of \thirteenCO, \CeighteenO, and other molecules
showing less pronounced peaks at the nucleus (No. 3) than at No. 2 and 4. 
This may accompany the possible higher temperature and hence stronger heating radiation at No. 3.

\subsection{Comparison with the Center of our Galaxy}
It is interesting to compare the central molecular zone of NGC 253 with that of our Galaxy
to learn the cause and consequence of starburst
because the two are (among the) nearest starburst and non-starburst galactic centers, respectively.
Both have near edge-on viewing angles and owe their gas distributions to a galactic stellar bar 
\citep{Scoville85,Sorai00,Binney91,Sawada04}. 
The Galactic CMZ is the central half kpc region of high gas concentration
with a total gas mass of \about $5\times10^{7}$ \Msol\ \citep{Pierce-Price00,Dahmen98}
and
a total  infrared ($\lambda =$ 1 --1000 \micron) luminosity from interstellar dust of 
$4\times 10^8$ \Lsol\ \citep{Sodroski97}.
Thus the starburst CMZ of NGC 253 has about an order of magnitude more gas
and is about two orders of magnitude more luminous than the central region of our Galaxy
(see references in \S \ref{s.intro}).  
A previous comparison of the two CMZs was made by \citet{Paglione95} in HCN(1--0) at \about50 pc resolution.

\begin{figure}
\epsscale{0.57}
\plotone{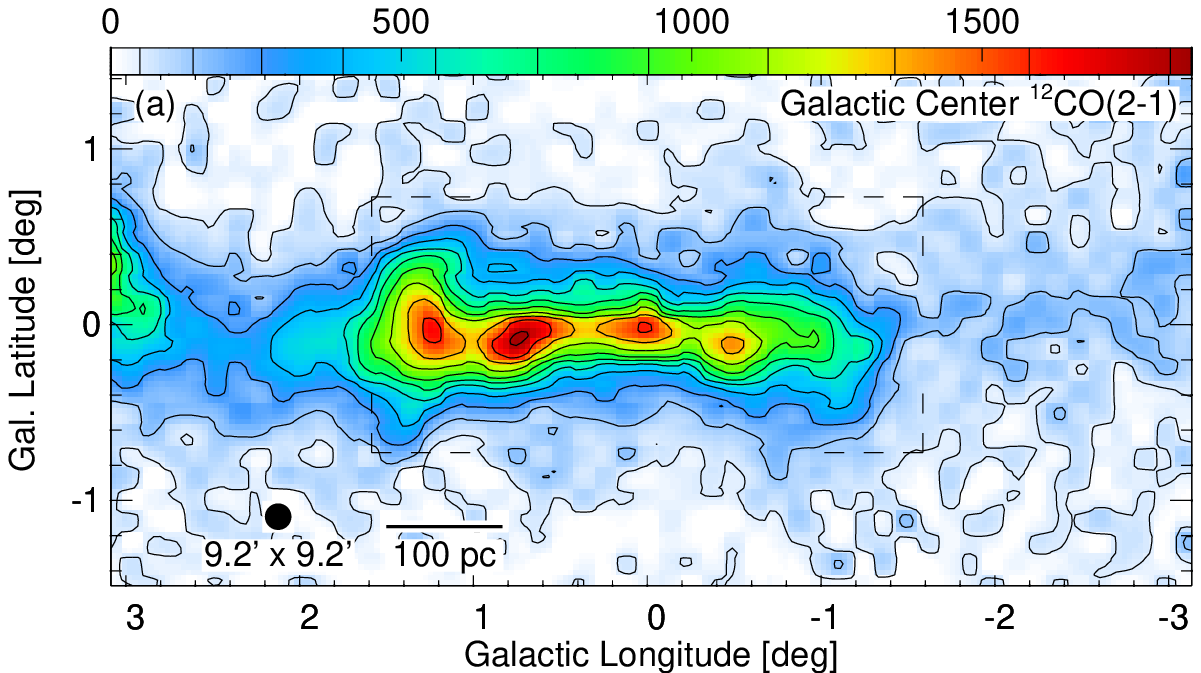} \\ 
\plotone{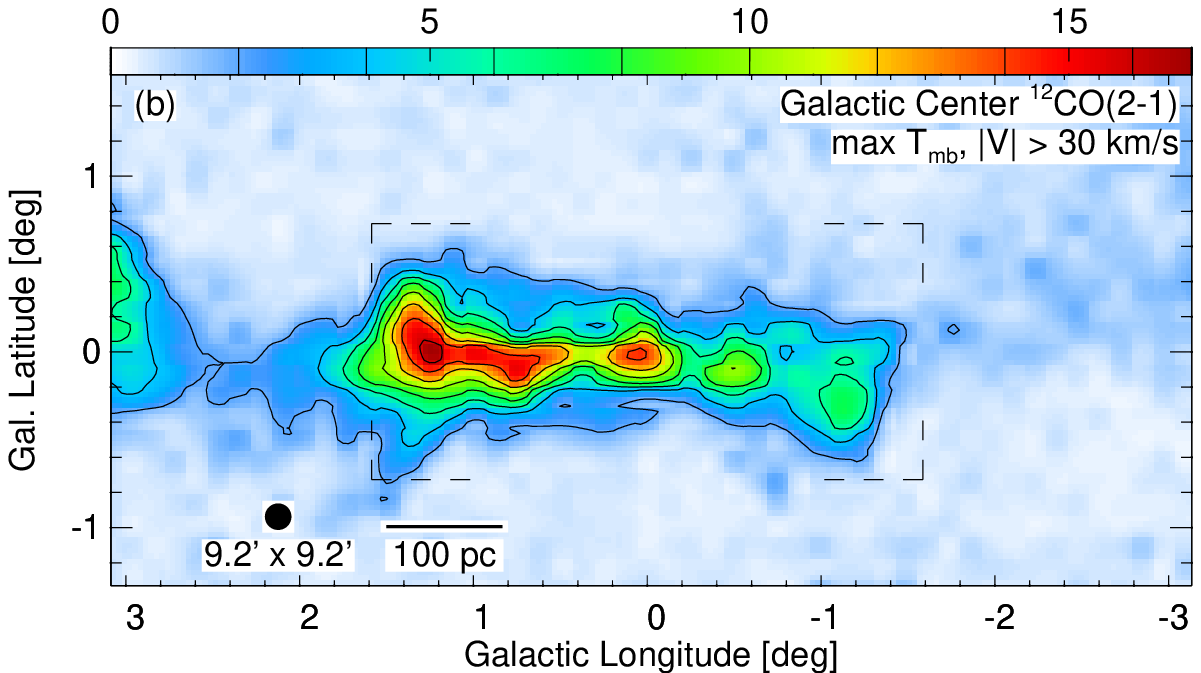} \\ 
\epsscale{1.0}
\caption{The center of the Milky Way galaxy in \twelveCO(2--1)
for comparison with the center of NGC 253. 
The single-dish data are from \citet{Sawada01} and have a resolution of
23 pc (FWHM) at the Galactic center (distance 8.5 kpc).
The dashed lines mark the same linear-scale area as they do in Fig. \ref{f.integ}.
(a) Integrated intensity. The contours are at $46\times n^{1.5}$ K(\Tmb) \kms.
The brightest peaks at $l=$ 0\arcdeg, 0.7\arcdeg, \minus0.5\arcdeg, and +1.3\arcdeg\ are
the Sgr A, B2, C clouds and the $l=1.3\arcdeg$ cloud, respectively.
(b) Peak brightness temperature with 2 K contour intervals.
Peak intensities were searched only at $\left| \Vlsr \right| > 30$ \kms\ 
to exclude emission from the disk and arms in the foreground or background of the Galactic center.
\label{f.gc} }
\end{figure}

We show \twelveCO(2--1) maps of the center of our Galaxy in Figure \ref{f.gc}
for comparison with NGC 253. 
The data were obtained by \citet{Sawada01} using the University of Tokyo 60 cm telescope
and have the same 20 pc resolution as our NGC 253 data. 
Two morphological similarities are evident from the comparison.
First, both galaxies have a similar size of about $0.5\times0.1$ \sqkpc\ at the
half maximum intensity of the CO(2--1) emission. 
The two central molecular zones therefore have about the same size.
Second, both galaxies show several prominent peaks of 20--50 pc sizes in their CMZs.
The four most prominent CO peaks in the Galactic CMZ are Sgr A, B2, and C cloud complexes
at $l=$ 0\arcdeg, 0.7\arcdeg, and \minus0.5\arcdeg, respectively, 
and the $l=1.3\arcdeg$ cloud complex.
About one third of the total CO(2--1) flux in the central 340 pc (FWHM) of our Galaxy is from the four
50 pc diameter regions centered at these molecular cloud complexes.
They show peaks in the map of peak brightness temperature (Figure \ref{f.gc}b).
They also have associated peaks of centimeter and (sub)millimeter continuum emission as in the case of NGC 253.
In particular, the Sgr B2 cloud complex contains the most
prominent sites of star formation in the Galactic center 
\citep[][and references therein]{Pierce-Price00,Schuller09,Bally10}.
The four cloud complexes in the Galactic center have
higher HCN-to-CO intensity ratios, or higher gas densities, than the area surrounding them according to the
$J$=1--0 line observations of the Galactic center by \citet{Jackson96}.
This coincides with what we saw in NGC 253 (\S \ref{s.line.ratio_density}) and constitutes 
the third element of similarities between the Galactic center and the center of NGC 253,
that is,
the massive molecular cloud complexes in both CMZs have high HCN to CO ratios probably due to their
high densities. 
This agrees with what \citet{Paglione95} found using 
HCN and CO (1--0) data of 3\asec\ and 7\asec\ resolution, respectively.

The largest difference between the two galaxies is that
the most prominent molecular cloud complexes in the center of NGC 253 have
a factor of 3--4 higher CO(2--1) intensities, in both integrated and peak values, than 
those in the center of our Galaxy.
The difference in the peak integrated intensities suggests, if the same  CO-to-\HH\ conversion factor applies to both,
that the cloud complexes in NGC 253 have higher masses and 
higher gas column densities at 20 pc scale than their Galactic counterparts.
Since the molecular cloud complexes in the two CMZs are similar in size according to our CO(2--1) maps, 
the higher column density in NGC 253 would also mean higher mean volume density 
in the cloud complexes in the starburst CMZ.
A caution here is that the uncertainty of the CO-to-\HH\ conversion factors in the centers of the two galaxies 
directly affects this comparison. 
We estimated in \S \ref{s.line.column_mass_opacity} the mass and mass surface densities of molecular gas 
in the center of NGC 253 using two methods independent of the conversion factor. 
Those estimates agree within a factor of a few to the mass derived from the adopted conversion factor of 
$\Xco = 1 \times 10^{20} $ \persquarecm\ (K \kms)$^{-1}$, suggesting the mass uncertainty of 0.5 dex.
The conversion factor in the center of our Galaxy also has its uncertainty.
For example, \citet{Dahmen98} mapped the Galactic center in optically thin \CeighteenO(1--0),
derived the mass of molecular gas, and compared the derived mass from other methods.
Their weighted best estimate for molecular mass in the central 600 pc
of our Galaxy is 4 times smaller than that from the \twelveCO(1--0) luminosity 
and the conversion factor we used for NGC 253. 
On the other hand,  \citet{Pierce-Price00} concluded from their Galactic center mapping in optically thin 
dust emission at 450 and 850 \micron\ 
that the total gas mass in the central 400 pc is 80\% larger than the best estimate mass
of \citet{Dahmen98} in the central 600 pc. If this were the correct mass the conversion factor
in the Galactic center is probably within a factor of 2 from what we adopted for NGC 253.
Further evaluation of the conversion factors in two galaxies (and their possible variation within each galaxy) 
is beyond the scope of this paper, but readers are cautioned about this uncertainty in 
translating the observed difference in CO integrated intensities to difference in mass surface densities.

The difference in the peak beam-averaged temperature is such that it is 50 K in NGC 253
and 17 K in the Galactic center at the same 20 pc resolution in the same CO(2--1) line.
The peak brightness temperature in the Galactic CMZ  is lower 
even at higher spatial resolution and with less beam dilution.
The peak \Tb\ of CO(1--0) and (3--2) emission 
in the Galactic CMZ is almost always below 30 K at 0.8 pc resolution \citep{Oka98,Oka07}.
Only in the Sgr A and B regions \citet{Oka01} found a cloud whose peak temperature exceeds 30 K;
the peak \Tb\ is 38 K and 31 K, respectively, at 2.5 pc resolution.
We therefore suggest that the difference in peak \Tb\ at 20 pc scale is mainly because 
the molecular complexes are warmer in NGC 253 than in the Galactic CMZ, 
although 50 K gas is not excluded in the Galactic CMZ at sub-parsec scales or behind the CO photospheres.
The warmer gas in NGC 253 is consistent with the larger luminosity of the starburst nucleus
and in accordance with the higher dust temperature there.
The dust temperature in the Galactic CMZ has been estimated to be \about20 K from
SED fitting in the submillimeter-to-far infrared bands \citep[][and references therein]{Pierce-Price00}
while dust in the center of NGC 253 was estimated also from SED fitting to be at or 
above 30 K \citep{Radovich01,Melo02,Weiss08}.
The high-temperature gas in the center of NGC 253 has also been pointed out from
the line excitation analyses mentioned in \S \ref{s.line.line_cont.comparison}.
Our \Tb\ observations add a new temperature constraint that does not 
involve such delicate issues in the line excitation modeling
as the assumption on cloud velocity structure.
Finally, the higher density and temperature of the molecular gas in the center of NGC 253
suggest that gas thermal pressure is also higher there
than in the Galactic CMZ, 
where \citet{Sawada01} estimated $P/k = \nhtwo T \sim 10^{5}$ \percubiccm\ K.
The difference will be about an order of magnitude on average if the
density increase in NGC 253 is in the microscopic $\nhtwo$.

\section{Summary \label{s.summary} }
The center of the nearby starburst galaxy NGC 253 has been imaged 
in five molecular lines and continuum at 0.87 mm and 1.3 mm at resolutions of 20--40 pc.  
A comparison was made in \twelveCO(2--1) with the center of our Galaxy at the same linear resolution.
Our main observations are the following.

1. Molecular gas in the center of NGC 253 is concentrated in an elongated region on the sky,
or a highly inclined disk, with the total extent of \about1 kpc, 
as has been seen in lower-resolution observations.
This region has a FWHM size of about 300 pc $\times$ 100 pc 
when observed in \twelveCO(2--1) at 20 pc resolution.
This size is approximately the same as the FWHM size of the Central Molecular Zone (CMZ)
of our Galaxy observed in the same line at the same linear resolution.

2. Within the CMZ of NGC 253, there is a string of peaks in molecular lines 
and 870--1300 \micron\  continuum emission. 
We reported the parameters of five most conspicuous peaks of $\lesssim$ (20--40) pc extent.
They are most likely molecular cloud complexes comparable in 
size with such molecular cloud complexes 
as Sgr A, B2, C and the $l=1.3\arcdeg$ complex in the CMZ of the Milky Way.

3. The molecular complexes in NGC 253 are warm, dense, and have high column densities.
They have 
\twelveCO\ peak brightness temperatures of \about50 K, 
molecular gas column density on the order of $10^{4}$ \Msol\ \persquarepc,
gas mass on the order of $10^{7}$ \Msol\ within a size scale of 20 pc,  
and volume-averaged gas density of $\nhtwo \sim 4000$ \percubiccm. 
HCN-to-CO ratio is higher at the molecular complexes than in the surrounding area,
as was observed in the Galactic CMZ, plausibly suggesting higher gas densities in the
cloud complexes.
There are indications from line and continuum ratios that the molecular complex
at the dynamical center has different properties, likely a higher gas temperature,
compared to other clouds.
Compared with their counterparts in the center of our Galaxy, the molecular complexes
in the CMZ of NGC 253 have peak brightness temperatures and column densities that are 
about a factor of 3 higher at the same spatial resolution, although the comparison of 
column density is subject to the uncertainty in the conversion factor.

4. The molecular complexes are spatially associated with the centimeter-wave radio peaks
such as those found by \citet{Turner85}. 
This coincidence, as well as the coincidence of the molecular complexes and other star formation
tracers, suggests that these molecular complexes are the sites of the most intense star formation
in the center of NGC 253. 
This is in parallel to the Sgr B2 cloud complex being the most
prominent site of star formation in the Galactic CMZ.

5. We find signs that the high opacity of \twelveCO\ prevents it from truly tracing gas distribution 
both in individual cloud complexes (because of self-absorption and cloud overlapping) 
and in the overall structure of the CMZ (because we only see the warm surface of the CO photosphere.)
Caution is needed about this bias when analyzing the gas parameters using 
only \twelveCO\ data.

\acknowledgements
We thank the referee for many constructive comments.
This research made use of 
the NASA/IPAC Extragalactic Database (NED),
NASA's Astrophysics Data System (ADS), 
and 
the computer system in the Astronomical Data Analysis Center (ADAC) of
the National Astronomical Observatory of Japan.
This work was partly supported by the National Science Council (NSC) grant 99-2112-M-001-011-MY3 to KS.
RQM was partly supported by the NSFC grants 10973042,10733030 and 10921063.

{\it Facilities:} \facility{SMA}


\clearpage 
\begin{deluxetable}{ccccccccc}
\tablewidth{0pt}
\tablecaption{Log of SMA observations  \label{t.obslog} }
\tablehead{ 
	\colhead{observation date}  &
	\colhead{main line} &
	\colhead{$f_{\rm LO}$ } &
	\colhead{$N_{\rm track}$ } &
	\colhead{$N_{\rm ant}$} &
	\colhead{$\tau_{225}$} &
	\colhead{$\langle T_{\rm sys, DSB}\rangle$} &
	\colhead{$L_{\rm baseline}$} &
	\colhead{$T_{\rm int}$} 
	\\
	\colhead{ }  & 
	\colhead{ }  &
	\colhead{GHz} &
	\colhead{ } &
	\colhead{ } &
	\colhead{} &
	\colhead{K} &
	\colhead{m} &
	\colhead{hr} 
	\\
	\colhead{(1)}  & 
	\colhead{(2)}  &
	\colhead{(3)} &
	\colhead{(4)} &
	\colhead{(5)} &
	\colhead{(6)} &
	\colhead{(7)} &
	\colhead{(8)} &
	\colhead{(9)} 
}
\startdata
2004 Aug.--2005 Nov. & CO(2--1)    & 224.650 &  6 &   7 or 8 &  0.07--0.12 & 130 & 8--509 & 5.7   \\
2004 Sep. & CO(3--2)    & 340.508 &  1 &   7 &  0.04 & 279 & 7--124 & 4.2   \\
2004 Nov. & HCN(4--3) & 349.222 &  1 &   7 &  0.09 & 447 & 7--124 & 3.3  
\enddata
\tablecomments{
(3) Frequency of the local oscillator.
The center frequencies of upper and lower sidebands are $f_{\rm  LO} \pm 5.0$ GHz.
Each sideband is 2 GHz wide.
(4) Number of tracks (i.e., nights) used. 
The 1.3 mm observations used array configurations called compact, compact-north, extended, and very-extended
for 1, 2, 2, and 1 night(s), respectively.
(5) Number of participating antennas.
(6) Zenith opacity at 225 GHz measured at the Caltech Submillimeter Observatory next to the SMA.
Opacity in the 350 GHz band is approximately $3.5\tau_{225}$.
(7) Median of double-side-band system temperature toward NGC 253.
(8) Range of projected baseline length for the galaxy.
(9) Total integration time on our target positions. The short $T_{\rm int}$ of our 230 GHz observations for the
number of tracks is because some off-center positions were also observed in these tracks. 
}
\end{deluxetable}

\begin{deluxetable}{cccclrrc}
\tablewidth{0pt}
\tablecaption{Data parameters  \label{t.dataparam} }
\tablehead{ 
	\colhead{line, band}       &
	\colhead{resolution} &
	\colhead{$\Delta v$} &
	\multicolumn{2}{c}{rms noise} &
	\colhead{$f_{\rm SD}$} &
	\colhead{$\theta_{\rm SD}$} &
	\colhead{SD ref.}
\\ 
	\colhead{ }       &
	\colhead{$\asec \times \asec$} &
	\colhead{\kms} &
	\colhead{mJy \perbeam} &
	\colhead{K} &
	\colhead{\%} &
	\colhead{\asec} &
	\colhead{}
}
\startdata
\twelveCO(2--1)       & $1.14\times1.13$ & 10  & 39    & 0.69   & 79 & 23 & 1\\
\thirteenCO(2--1)     & $1.62\times1.49$ & 20  & 17    & 0.18   & 89 & 23 & 1, 2\\
\CeighteenO(2--1)   & $1.62\times1.49$ & 20  & 18    & 0.19   & 97 & 23 & 1\\
1.33 mm continuum   & $1.10\times1.10$ & \nd & 2.5   & 0.050 & $\geq$64 & \about30 & 3, 4\\
\twelveCO(3--2)        & $1.82\times1.10$ & 10  & 215  & 1.2     & 62 & 22 & 5\\
HCN(4--3)                  & $2.20\times1.71$ & 30  & 130 & 0.34   & 72 & 20 & 6\\
0.87 mm continuum & $1.80\times1.10$ & \nd & 23    & 0.12   & 60 & 15 & 7, 8\\
\enddata
\tablecomments{The resolutions are in full width at half maximum (FWHM). 1\asec\ = 17 pc. 
The last three columns are 
fractions of single dish (SD) fluxes detected in our data, 
the FWHM beam sizes of the single dish observations, 
and the sources of the single dish data.
The comparisons are made at the center positions given in the single dish papers.
Line contamination in the single-dish continuum observations has been subtracted adopting
the estimates of \twelveCO\ contamination by \citet{Mauersberger96} and \citet{Seaquist04}; 
specifically the subtraction was 25\% and 30\% for the 1.33 and 0.87 mm, respectively.
The uncertainty of $f_{\rm SD}$ is about 15\% and 20\% for 1.3 mm and 0.8 mm 
data, respectively. 
}
\tablerefs{
1. \citet{Harrison99};
2. \citet{Israel02};
3. \citet{Krugel90}; 
4. \citet{Mauersberger96};
5. \citet{Bayet04};
6. \citet{Knudsen07};
7. \citet{Alton99};
8. \citet{Seaquist04}
}
\end{deluxetable}

\begin{deluxetable}{ccccccccccl}
\tablewidth{0pt}
\tablecaption{Near 1 mm Continuum Peaks in the Center of NGC 253  \label{t.cont_peaks} }
\tablehead{ 
	\colhead{No.}       &
	\colhead{$\alpha$} &
	\colhead{$\delta$} &
	\colhead{$I_{\rm 1.3}$} &
	\colhead{$I_{\rm 0.87}$} &
	\colhead{$R_{0.87/1.3}$} &
	\multicolumn{2}{c}{$\Sigma_{g}$} &
	\multicolumn{2}{c}{$M_{g}$} &				
	\colhead{association}	
\\
	\colhead{ }       &
	\colhead{(00\hr47\mn)} &
	\colhead{(\minus25\arcdeg17\arcmin)} &
	\multicolumn{2}{c}{Jy \perbeam} &
	\colhead{} &
	\multicolumn{2}{c}{$10^4$ \Msol\ \persquarepc} &
	\multicolumn{2}{c}{$10^7$ \Msol} &				
	\colhead{ }	
\\
	\colhead{(1)} &
	\colhead{(2)} &
	\colhead{(3)} &
	\colhead{(4)} &
	\colhead{(5)} &
	\colhead{(6)} &	
	\colhead{(7)} &
	\colhead{(8)} &
	\colhead{(9)} &
	\colhead{(10)} &
	\colhead{(11)} 	
}
\startdata
1 & 33\fs63 & 13\farcs2 & 
0.017 &  (0.04) & \nd & 
0.9 & (0.2) & 0.4 & (0.2) &
\water\_2 
\\
2 & 33\fs30 & 15\farcs6 & 
0.064 & (0.22) & 2.7 &
3.5 & (1.3) & 1.5 & (0.9) &
TH1 
\\
3 & 33\fs17 & 17\farcs1 & 
0.082 & 0.33 & 3.0 &
4.5 & 2.0 & 1.9 & 1.3 &
TH2$\approx$nucleus, TH4=\water\_1, TH3,5,6
\\
4 & 32\fs99 & 19\farcs7 & 
0.063 & 0.27 & 3.6 &
3.4 & 1.6 & 1.4 & 1.0 &
TH7(=SSC),TH8
\\
5 & 32\fs84 & 21\farcs3 & 
0.044 & 0.21 & 3.5 &
2.4 & 1.3 & 1.0 & 0.8 &
TH9
\enddata
\tablecomments{ 
Columns 2 and 3 :  
Peak position of 1.3 mm continuum in J2000. 
Seconds of right ascension and arcseconds of declination, respectively.
Columns 4 and 5 : 
Peak flux density of 1.3 mm and 0.87 mm continuum, respectively.
The conversion factor to brightness temperature is 19.9 K/(Jy \perbeam) for 1.3 mm
and 5.2 K/(Jy \perbeam) for 0.87 mm.
The 0.87 mm intensities in parenthesis are measured at the corresponding 1.3 mm peak positions
because 0.87 mm data lack an isolated peak.
Column 6 :  
Ratio of 0.87 mm to 1.3 mm continuum flux densities measured in the data with matching \uv\ coverage
and 2\asec\ resolution. The fractional error of the ratio is 20\%. 
The peak No. 1 was too weak at 0.87 mm to reliably derive the ratio.
Columns 7 and 8 : 
Peak gas surface density estimated from 1.3 mm and 0.87 mm data, respectively,
with $\Sigma_g = 5.4\times 10^5 \times(I_{1.3}/\mbox{Jy beam$^{-1}$})$ \Msol\ \persquarepc\ 
and $\Sigma_g = 6.0\times 10^4 \times(I_{0.87}/\mbox{Jy beam$^{-1}$})$ \Msol\ \persquarepc.
See \S \ref{s.continuum.mass} for assumptions and uncertainties.
Columns 9 and 10 : 
Gas mass of each peak in the 1.3 mm and 0.87 mm beam, respectively.
The beam FWHM is $19 \times 19$ pc$^2$ and $31 \times 19$ pc$^2$ in our 1.3 mm and 0.87 mm data,
respectively.
Column 11 : 
Association with sources known in other wavelengths.
Compact centimeter sources of \citet{Turner85} are called TH$n$, among which
TH2 is thought to be at the galactic center \citep[or within 2\arcsec of it;][]{Fernandez-Ontiveros09}.
The nuclear region has two water masers \water\_1 (at $\Vlsr \approx 120$ \kms) and  \water\_2  (170 \kms) 
associated with 
star formation \citep{Henkel04,Hofner06,Brunthaler09}. 
\water\_2 is likely associated with the molecular cloud complex that has the peak line velocity of 170 \kms\ (see
Fig. \ref{f.ispec5in1}) and includes the continuum peak No. 1 listed above.
SSC is the most luminous mid-IR source in the nucleus and is suggested to be a super star cluster
\citep{Keto99}. 
}
\end{deluxetable}

\begin{deluxetable}{cccccccccccccccc}
\tablewidth{0pt}
\tablecaption{Molecular-cloud Complexes in the Center of NGC 253  \label{t.line_peaks} }
\tablehead{ 
	\colhead{}       &
	\colhead{00\hr47\mn} &
	\colhead{\minus25\arcdeg17\arcmin} &	
	\multicolumn{3}{c}{CO(2--1)} &
	\multicolumn{2}{c}{CO(3--2)} &	
	\multicolumn{3}{c}{\thirteenCO(2--1)} &
	\multicolumn{3}{c}{\CeighteenO(2--1)} &
	\multicolumn{2}{c}{HCN(4--3)} 
\\
	\colhead{No.}       &
	\colhead{$\alpha$} &
	\colhead{$\delta$} &
	\colhead{\Idv} &
	\colhead{\Tp} &
	\colhead{\Ipdv} & 	
	\colhead{\Idv} & 
	\colhead{\Tp} &
	\colhead{\Idv} & 
	\colhead{\Tp} &
	\colhead{\taubar} &	
	\colhead{\Idv} & 
	\colhead{\Tp} &	
	\colhead{\taubar} &
	\colhead{\Idv} & 
	\colhead{\Tp} 			
\\
	\colhead{(1)} &
	\colhead{(2)} &
	\colhead{(3)} &
	\colhead{(4)} &
	\colhead{(5)} &
	\colhead{(6)} &
	\colhead{(7)} &
	\colhead{(8)} &	
	\colhead{(9)} &
	\colhead{(10)} &
	\colhead{(11)} &
	\colhead{(12)} &
	\colhead{(13)} &
	\colhead{(14)} &
	\colhead{(15)} &
	\colhead{(16)} 		
}
\startdata
1 & 33\fs67 & 13\farcs1 & 
5.5e3 & 52 & 4.9e3 &
4.2e3 & 41 & 
5.1e2 & 6.0 & 0.11 & 
1.4e2 & 2.2 & 0.028 &
2.6e2 & 2.9 
\\
2 & 33\fs38 & 15\farcs8 & 
6.1e3 & 43 & 5.3e3 &
4.5e3 & 36 & 
8.5e2 & 10.4 & 0.17 & 
2.8e2 & 4.7 & 0.054 &
3.7e2 & 3.6 
\\
3 & 33\fs18 & 17\farcs4 & 
6.5e3  & 34 & 5.6e3 &
5.7e3 & 34 & 
7.3e2 & 5.1 & 0.14 &
2.2e2 & 1.5 & 0.040 &
4.6e2 & 2.5 
\\
4 & 32\fs99 & 19\farcs6 & 
6.2e3 & 48 & 5.6e3 &
5.2e3 & 40 & 
8.4e2 & 7.7 & 0.16 &
2.3e2 & 2.8 & 0.042 &
4.1e2 & 5.1
\\
5 & 32\fs82 & 20\farcs9 & 
5.0e3 & 47 & 4.6e3  &
3.7e3 & 36 & 
7.9e2 & 10.5 & 0.19 &
2.1e2 & 4.5 & 0.047 &
2.6e2 & 3.0 
\\
\enddata
\tablecomments{ 
Columns 2 and 3 :  
Peak positions of \twelveCO(2--1) integrated intensity, except for peak No. 5 whose position is from 
the \thirteenCO(2--1) integrated map because \twelveCO(2--1) does not show a peak.  
Seconds of right ascension and arcseconds of declination, respectively, in J2000.
Columns 4, 7, 9, 12, 15 :  
\Idv\ is the integrated intensity in K \kms\ at the position of (2) and (3). 
Columns 5, 8, 10, 13, 16 :  
\Tp\ is the peak intensity in K at the position of (2) and (3).
Both \Idv\ and \Tp\ are measured from the data at their native spatial resolutions in Table \ref{t.dataparam}.
Column 6 : 
\Ipdv\ is the integrated intensity in K \kms\ at the position of (2) and (3) measured after the data 
are convolved to the spatial resolution of the \thirteenCO(2--1) and \CeighteenO(2--1) data.
Columns 11 and 14 : 
Mean opacity of the line, \taubar, estimated from the ratio of integrated intensities
between the isotopologue and \twelveCO.
}
\end{deluxetable}


\clearpage

\end{document}